\documentclass[prb,aps,showpacs]{revtex4}

\usepackage{graphicx}

\usepackage{amsmath}

\usepackage{amssymb}

\usepackage{epsfig}

\begin{document}

\newcommand{\be}{\begin{equation}}
\newcommand{\ee}{\end{equation}}



\title{Scaling properties of step bunches induced
by sublimation and related mechanisms:  
A unified perspective}

\author{J. Krug}
\affiliation{Institut f\"ur Theoretische Physik, Universit\"at zu K\"oln,
Z\"ulpicher Strasse 77, 50937 K\"oln, Germany}

\author{V. Tonchev, S. Stoyanov}
\affiliation{Institute of Physical Chemistry, Bulgarian Academy of Sciences,
1113 Sofia, Bulgaria}

\author{A. Pimpinelli}
\affiliation{LASMEA, UMR 6602 CNRS/Universit\'e Blaise-Pascal -- Clermont 2, 
F-63177 Aubi\'ere Cedex, France}

\date{\today}

\begin{abstract}
This work provides a ground for a quantitative interpretation of
experiments on step bunching during sublimation of crystals with a
pronounced Ehrlich-Schwoebel (ES) barrier in the regime
of weak desorption. A strong step bunching instability
takes place when the kinetic length
$d_d = D_s/K_d$ is larger than the average distance $l$
between the steps on the vicinal surface; here $D_s$ is the surface diffusion
coefficient and $K_d$ is the step kinetic coefficient. 
In the opposite limit $d_d \ll l$ the instability is weak and step
bunching can occur only when the magnitude of step-step repulsion
is small. The central result are 
power law relations of the form $L \sim H^\alpha$, $
l_{\mathrm{min}} \sim H^{-\gamma}$ between the width $L$, the height $H$, and
the minimum interstep distance $l_{\mathrm{min}}$ of a bunch.
These relations are obtained from a continuum evolution
equation for the surface profile, which is derived from
the discrete step dynamical equations for the case $d_d \gg l$. 
The analysis of the continuum equation reveals the existence of
two types of stationary bunch profiles with different scaling
properties. Through comparison with numerical simulations of the discrete
step equations, we establish the value $\gamma = 2/(n+1)$ for the 
scaling exponent of $l_{\mathrm{min}}$ in terms
of the exponent $n$ of the repulsive step-step interaction,
and provide an exact expression
for the prefactor in terms
of the energetic and kinetic parameters of the system.
For the bunch width $L$ we observe significant deviations
from the expected scaling with exponent  
$\gamma = 1 - 1/\alpha$, which are attributed to the pronounced
asymmetry between the leading and the trailing edges of the bunch, and
the fact that bunches move. Through a mathematical equivalence on the
level of the discrete step equations as well as on the continuum level,
our results carry over to the problems of step bunching induced
by growth with a strong inverse ES effect, and by electromigration
in the attachment/detachment limited regime. Thus our work provides
support for the existence of universality
classes of step bunching instabilities [A. Pimpinelli et al., Phys. Rev. Lett.
\textbf{88}, 206103 (2002)], but some aspects of the universality
scenario need to be revised.  

\end{abstract}


\maketitle

\section{Introduction}
\label{Introduction}

The formation of step bunches at a vicinal surface
is a problem of great current interest, both from a fundamental
viewpoint and with regard to the possible uses of step bunches as nanotemplates
or nanostructures \cite{Venezuela99,Teichert02,Neel03,Syvajarvi02,Gliko02,Gliko03}. 
Mechanisms causing step bunching instabilities 
include strain effects \cite{Venezuela99,Teichert02,Tersoff95,Liu98}, sublimation under
conditions of asymmetric detachment kinetics known as the Ehrlich-Schwoebel (ES)
effect \cite{Schwoebel69,Pimpinelli94,Uwaha95}, 
growth with an inverse ES effect \cite{Schwoebel69,Sato01,Xie02,Tonchev03},
and surface electromigration
\cite{Latyshev89,Homma90,Stoyanov91,Houchmandzadeh94,Yang96,Dobbs96,Stoyanov97,Stoyanov97a,Fu97,Liu98a,Metois99,Sato99,Yagi01,Minoda03}.  

Quite recently
it was realised that step bunching is a promising way to study the
interactions between the steps \cite{Stoyanov98a,Stoyanov98b,Fujita99,Homma00,Stoyanov00,Stoyanov00a}. 
The physical ground is simple: The steps in the bunch
keep a certain distance from each other because the step-step repulsion
balances the tendency to further compression of the bunch. The free
energy related to the step-step interaction is of the form $A/l^n$,
where $l$ is the interstep distance. When $n=2$, the amplitude $A(T)$ accounts for both
elastic and entropic repulsion between the steps \cite{Nozieres91}. Under crystal-vapour
equilibrium one has the relation $A(T) \sim g(T)$ where $g(T)$ is the step
repulsion coefficient in the expression
\be
f(\rho) = f(0) + \kappa\rho + g\rho^3                           \label{Gl1}
\ee
for the surface free energy (per unit projected area) of a vicinal crystal
surface with a density of steps $\rho$. 

\begin{figure}
\centerline{\epsfig{file=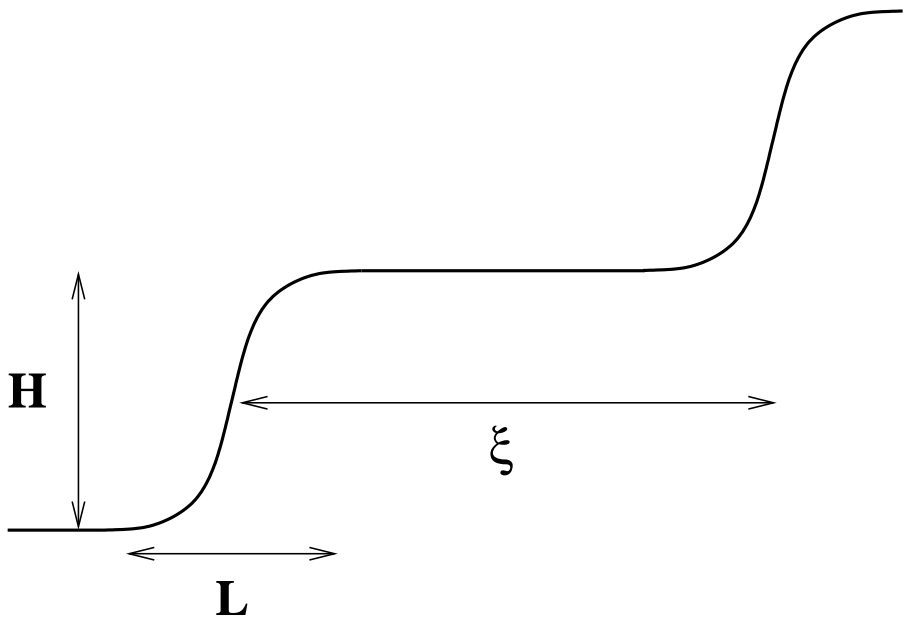,width=12.cm}} 
\vspace*{0.5cm}
\caption{\label{Fig0} Schematic of a bunched vicinal surface, illustrating the definition 
of the bunch width $L$, the bunch height $H$ and the bunch spacing $\xi$.}
\end{figure}

To infer information about the step-step interactions from experimental
observations of bunch morphology, one makes use of scaling relations
between the length and time scales characterizing the bunches. 
The relevant length scales are the width $L$ and the height $H$ of the 
bunch, and the spacing $\xi$ between subsequent bunches (Fig.\ref{Fig0}).
The length $\xi$ is also sometimes referred to as the terrace width;
this nomenclature is somewhat ambiguous, however, because the region
between two bunches may contain several monoatomic steps, and, hence,
several wide terraces.  
The bunch height is related to the number of steps $N$ in the bunch
by $H = N h_0$, where $h_0$ is the height of an atomic step.
A quantity that is directly accessible to experimental observations
\cite{Fujita99} is the minimal terrace size $l_{\mathrm{min}}$ inside
the bunch, which is related to the maximal slope $m_{\mathrm{max}}$
through $l_{\mathrm{min}} = h_0/m_{\mathrm{max}}$. Following the notation of 
Ref.36,  
we introduce scaling exponents $\alpha$ and $\gamma$ characterizing 
the shape of individual bunches through the relations
\be
\label{StatExp}
H \sim L^\alpha, \;\;\; l_{\mathrm{min}} \sim N^{-\gamma}.
\ee
Assuming that the minimal terrace size
$l_{\mathrm{min}}$ is of the same order as the mean size
$\bar l = L/N$ of the terraces inside the bunch, one expects 
the exponent identity 
\be
\label{scalingrelation}
\gamma = 1 - 1/\alpha.
\ee
Furthermore the coarsening of the bunch morphology with time is described
by a dynamic exponent $z$ defined through \cite{Pimpinelli02} 
\be
\label{DynExp}
\xi \sim N \sim t^{\alpha/z}.
\ee
Because the ratio $\xi/N$ has to be equal to the mean terrace
width $l$, which is fixed by the overall vicinality of the surface, 
the bunch spacing $\xi$ and the bunch size $N$
grow with time in the same manner.

For step bunching induced by surface electromigration, 
scaling relations of the form (\ref{StatExp}) have been derived 
both for non-transparent and transparent steps \cite{Sato99,Stoyanov98a,Stoyanov98b,Stoyanov00a}.
Their application to the analysis of experiments \cite{Fujita99,Homma00} proceeds in two stages.
First, the value of the scaling exponent $\alpha$ or $\gamma$ is used to determine
the kinetic regime and the value of the step interaction exponent $n$, and subsequently
an estimate for the strength of the step-step interaction (in relation to the driving force
for step bunching) is extracted from the prefactor of the scaling relation. For example,
for non-transparent steps one finds that $\alpha = (n+1)/(n-1)$, which yields
$\alpha = 3$ in the standard case $n=2$, whereas for $n=1$ the width of the bunch $L$
would in fact be independent of the bunch size.  
Provided that $n=2$ and $\alpha = 3$, the bunch width depends on the
step interaction coefficient as $L \sim [A(T)]^{1/3}$. This
relation provides a ground to study the temperature dependence of
$A(T)$ by measuring the width of the bunch as a function of temperature \cite{Stoyanov00}.

The main purpose of this paper to derive scaling relations of the form
(\ref{StatExp}) for step bunches induced 
by the ES effect during sublimation, and to put the results into 
perspective with regard to previous work on
other step bunching instabilities.  
In the next section we introduce the basic
concepts and quantities of the Burton, Cabrera, Frank (BCF) model 
\cite{Burton51} in the presence of ES barriers (see Ref.38 for a recent review). 
In Sect.\ref{Equationsofmotion}
the equations of step motion are displayed and various limiting cases
are discussed. In Sections \ref{Contrelax} and \ref{Continuum} 
a continuum evolution equation for the surface is derived, and the
structure of stationary bunch solutions is analysed. A key result is
the existence of two types of solutions with \textit{different}
scaling properties in the sense of (\ref{StatExp}). 
In Sect.\ref{EquivalentProblems} the mathematical equivalence of the sublimation
problem to appropriate limiting cases of step bunching induced by electromigration
and growth with inverse ES barriers is pointed out and exploited. 
Section\ref{Numerics} presents results obtained from 
numerical simulations of the discrete step dynamics and compares them to 
the predictions of the continuum theory. 
In Sect.\ref{Universality} we critically examine \cite{Krug04a} a recently proposed
classification scheme for step bunching instabilities \cite{Pimpinelli02}
in the light of our results, and some general conclusions
are drawn in Sect.\ref{Conclusions}. 

\section{Discrete model and basic concepts of sublimation by step
         flow}

We consider a vicinal surface going uphill in the $+ x$ direction.
The processes of atom migration (in the presence of desorption and
deposition) are described by the stationary diffusion equation

\be
D_s \frac{d^2n_s}{dx^2} - \frac{n_s}{\tau_s} + R = 0            \label{Gl2}
\ee
where $D_s$ is the coefficient of surface diffusion, $n_s$ is the
concentration of mobile atoms, adsorbed on the surface, $R$ is the
deposition rate of atoms to the crystal surface, $\tau_s$ is the life
time of an atom in a state of mobile adsorption, and $x$ is a coordinate
perpendicular to the step edges (we consider a system of parallel steps with
straight edges). The exchange of atoms between
the crystal phase and the dilute layer of adatoms takes place at the
steps and determines the boundary conditions for Eq.~(\ref{Gl2}).
It is essential to note that the ES barrier
considerably decreases the permeability of the steps. Really, an
atom has to break many chemical bonds when it crosses over the
step down to a position of adsorption at the step edge. Therefore,
such an event happens very rarely. The opposite jump from the position
of adsorption at the step edge to a position of adsorption on the
upper terrace (or, briefly, from the lower to the upper terrace)
also happens rarely in accordance to the principle of the detailed
balance. This circumstance justifies the reduction of the diffusion
problem on the crystal surface to a diffusion problem at a single
terrace. The boundary conditions relate the surface fluxes of
adatoms with the power of the step as a generator (or a sink) of
adatoms. For the terrace between the $i$-th and $i + 1$-th step
one can write

\be
- D_s \frac{dn_s}{dx}\vert_{x=x_i} =
-K_u [n_s(x_i) - n^e_s(x_i)]                     \label{Gl3} 
\ee
$$
- D_s \frac{dn_s}{dx} \vert_{x=x_{i+1}} = 
K_d [n_s(x_{i+1}) - n^e_s(x_{i+1})]           
$$
where $K_d$ and $K_u$ are the kinetic coefficients for ascending and
descending steps. The step kinetic coefficients $K_d$ and $K_u$ are
defined by the expression for the velocity of the step motion

\be
v = \frac{dx_i}{dt} = - a_{\perp} a_\parallel \{ 
      K_d [n_{s,i-1}(x_i) - n^e_s(x_i)]
        + K_u [n_{s,i}(x_i) - n^e_s(x_i) ] \}                        \label{Gl4}
\ee
where $a_{\perp}$ and $a_\parallel$ are the interatomic distances
perpendicular and parallel to the steps, $n_{s,i-1}(x)$ is the adatom
concentration on the terrace between the $i - 1$-th and the $i$-th step, and
$n^e_s(x_i)$ is the equilibrium value of the adatom concentration in
the vicinity of the step, situated at $x_i$. The value of $n^e_s$ in the
vicinity of the $i$-th step depends on the distances to the neighboring steps
$i+1$ and $i-1$, since 
$n^e_s(x_i) = n^e_s \exp[\Delta\mu(x_i)/k_{\mathrm{B}}T]$, where
\cite{Stoyanov98a,Stoyanov98b}
\be
\frac{\Delta\mu(x_i)}{k_{\mathrm{B}}T} 
= -\ \left(\frac{l_0}{x_{i+1}-x_{i}}\right)^3 +
        \left(\frac{l_0}{x_i - x_{i-1}}\right)^3
                                                                \label{Gl5}
\ee

and

\be
l_0 = \left(\frac{2 \Omega A(T)}{k_{\mathrm{B}}T}\right)^{1/3}
                                                                \label{Gl6}
\ee
is a length scale characterizing the strength of the step-step interaction.
In (\ref{Gl6}) we have introduced the atomic area $\Omega = a_\perp a_\parallel$.

The Ehrlich-Schwoebel barrier \cite{Michely03} provides the physical ground for the
inequality $K_u \neq K_d$, since the exchange of atoms between the crystal
phase and the adlayer on the upper terrace has a lower rate than the
exchange between the crystal and the adlayer at the lower terrace,
i.e., $K_u < K_d$, although the opposite inequality has also been
discussed \cite{Chung02}. 
Assuming that the step kinetic coefficients can be written
in the form $K_{u,d} = K_0 \exp ( - E_{u,d}/k_{\mathrm{B}}T)$ one
can write $K_u/K_d = \exp(- \Delta E/k_{\mathrm{B}}T)$
where $\Delta E = E_u - E_d$ characterises the asymmetry in the atom
attachment-detachment kinetics at the steps. The ratio
$\beta = K_u/K_d$ is an essential parameter in the considerations
presented below.

\section{Equations of step motion}

\label{Equationsofmotion}

Since the normal ES barrier $(\Delta E = E_u - E_d > 0)$ causes a
step bunching instability only in the case of sublimation \cite{Schwoebel69}, the equations
of step motion will be derived under the condition $R = 0$ (i.~e., in the
absence of deposition). The steps then move to the
right. Solving the diffusion problem [Eq.~(\ref{Gl2})
with the boundary conditions (\ref{Gl3})] one obtains an expression for the
adatom concentration at the crystal surface. Substituting $n_{s,i-1}(x_i)$
and $n_{s,i}(x_i)$ into Eq.~(\ref{Gl4}) one can write an expression
for the velocity $dx_i/dt$ of the $i$-th step as a function of the
widths of the lower (left) and upper (right) terraces 
\be
\frac{dx_i}{dt} = v_u + v_d.                                 \label{Gl7}
\ee
In the physically interesting limit of small desorption rate, in the sense
that the diffusion length $\lambda_s = \sqrt{D_s \tau_s} \gg x_{i+1} - x_i$,
the two contributions to the step velocity (\ref{Gl7}) read
\be
v_u = \frac{D_s \Omega n^e_s}{\lambda_s} 
 \frac{\beta \left[\frac{\Delta\mu(x_i)}{k_{\mathrm{B}}T} - 
  \frac{\Delta\mu(x_{i+1})}{k_{\mathrm{B}}T}\right]  + 
    \frac{\beta d_d}{\lambda_s}
     \left(\frac{x_{i+1}-x_{i}}{\lambda_s}\right) +
     \beta \left(\frac{x_{i+1}-x_{i}}{\lambda_s}\right)^2}
       {\frac{d_d}{\lambda_s} (1 + \beta) + 
        \left[\beta + \left(\frac{d_d}{\lambda_s}\right)^2\right]
        \left(\frac{x_{i+1}-x_{i}}{\lambda_s}\right)}             \label{Gl8a}
\ee
and 
\be
v_d = \frac{D_s \Omega n^e_s}{\lambda_s} 
 \frac{\beta \left[\frac{\Delta\mu(x_i)}{k_{\mathrm{B}}T} - 
  \frac{\Delta\mu(x_{i-1})}{k_{\mathrm{B}}T}\right] + 
    \frac{d_d}{\lambda_s}
     \left(\frac{x_i-x_{i-1}}{\lambda_s}\right) +
     \beta \left(\frac{x_i-x_{i-1}}{\lambda_s}\right)^2}
       {\frac{d_d}{\lambda_s} (1 + \beta) + 
        \left[\beta + \left(\frac{d_d}{\lambda_s}\right)^2\right]
        \left(\frac{x_i-x_{i-1}}{\lambda_s}\right)}             \label{Gl8b}
\ee
Here we have introduced the kinetic lengths \cite{Pimpinelli94,Krug04b}
$d_d = D_s/K_d$ and
$d_u = D_s/K_u = D_s/K_d\beta = d_d/\beta$.
The equations (\ref{Gl8a},\ref{Gl8b}) have two limiting cases:
\be
{\mathrm{(a)}} \;\;\; \frac{d_d}{\lambda_s} (1 + \beta)  \ll 
  \left[\beta + \left(\frac{d_d}{\lambda_s}\right)^2\right]
   \left(\frac{x_i - x_{i-1}}{\lambda_s}\right)  
\label{Gl9a}                         
\ee
\be
{\mathrm{(b)}} \;\;\; \frac{d_d}{\lambda_s} (1 + \beta)  \gg 
\left[\beta + \left(\frac{d_d}{\lambda_s}\right)^2\right]
   \left(\frac{x_i - x_{i-1}}{\lambda_s}\right)                 
\label{Gl9b}
\ee
The limit (a) (realised when $x_i - x_{i-1} \gg d_d$
and $d_d/\lambda_s \ll 1$) reduces the denominators
of Eqs.\ (\ref{Gl8a},\ref{Gl8b}) to $(\beta/\lambda_s)(x_i - x_{i-1})$. If in
addition $\beta \gg (d_d/\lambda_s)^2$, the 
parameter $\beta$ is eliminated from the terrace-length dependent
destabilizing terms in  
(\ref{Gl8a}) and (\ref{Gl8b}), and appears only in a constant contribution
to the step velocity, where it
does not provide any ground for a step bunching instability.
It is, however, quite possible that an instability is induced by higher order
terms $[(x_i - x_{i-1})/\lambda_s]^\nu$ with $\nu > 2$.
We shall address this possibility in Sect.\ref{Numerics}, devoted to numerical
analysis of the discrete model.

The limit (b) is more interesting. It takes place under the assumption
$x_i - x_{i-1} \ll d_d$ and $d_d/\lambda_s < 1$. Then only the constant
terms in the denominator of Eqs.(\ref{Gl8a},\ref{Gl8b}) are retained and
the terms quadratic in the terrace lengths can be neglected relative to the linear
ones. Thus Eqs.\ (\ref{Gl8a},\ref{Gl8b}) reduce to
\be
v_u  =  \frac{D_s \Omega n^e_s}{d_d (1 + \beta)}\ 
  \left\{ \frac{\beta}{k_{\mathrm{B}}T} [\Delta\mu(x_i) - 
  \Delta\mu(x_{i+1})]\  + 
    \frac{\beta d_d}{\lambda_s^2} (x_{i+1}-x_i) \right\}
\label{Gl10a}              
\ee
\be
v_d  =  \frac{D_s \Omega n^e_s}{d_d (1 + \beta)}\ 
  \left\{ \frac{\beta}{k_{\mathrm{B}}T} [\Delta\mu(x_i) - 
  \Delta\mu(x_{i-1})]\  + 
    \frac{d_d}{\lambda_s^2} (x_{i}-x_{i-1}) \right\}.
\label{Gl10b}                      
\ee
The two terms in the curly brackets have a clear physical
meaning. The first term (the difference between the chemical potentials
of neighbouring steps) is the driving force for the relaxation of the
fluctuations in the step density. The second term in Eqs.\ 
(\ref{Gl10a},\ref{Gl10b})
reflects the asymmetry in the step kinetics and provides a ground for
step bunching instability. These two terms describe a model 
previously analysed in Ref.42, where it was shown how continuum equations
for the one-dimensional crystal profile $h(x,t)$ can be derived exactly,
provided the step dynamics is linear in the terrace lengths. Indeed,
apart from the chemical potential difference in the square
brackets, we can write
\be
\frac{dx_i}{dt} = 
 v_u + v_d = \frac{D_s \Omega n^e_s}{\lambda^2_s}\  
  \left[\frac{\beta}{1 + \beta}\ (x_{i+1} - x_i) + 
   \frac{1}{1+\beta} (x_i - x_{i-1})\right]                     \label{Gl11}
\ee
where 
\be
\label{Rhat}
\frac{D_s \Omega n^e_s}{\lambda^2_s}  =
\frac{\Omega  n^e_s}{\tau_s} = 
\Omega R_e \equiv \hat R_e
\ee
is the equilibrium value of the
desorption rate per adsorption site. We will return to the continuum equation
derived from (\ref{Gl11}) below in Sect.\ref{Continuum}. 

\section{Continuum limit of the relaxational dynamics}
\label{Contrelax}

In this section we develop a continuum description for the relaxational part
of the step dynamics. 
The chemical potential differences in Eqs.\ (\ref{Gl10a},\ref{Gl10b}) can be written
approximately in the form 
\be
\frac{1}{k_{\mathrm{B}}T} \left[\Delta\mu(x_i) - 
  \Delta\mu(x_{i-1}) \right] \approx \frac{1}{k_{\mathrm{B}}T}
   \left[\frac{\partial}{\partial x}\ (\Delta\mu)\right]\ (x_i - x_{i-1}).
                                                                \label{Gl12}          
\ee
Making use of the relation $(x_i - x_{i-1}) \approx
h_0 (\partial h/\partial x)^{-1}$, one can bring the last expression into the
form
\be
\frac{h_0}{k_{\mathrm{B}}T} \left( \frac{\partial h}{\partial x} \right)^{-1}\ 
 \left[\frac{\partial}{\partial x}\ \bigg(\Delta\mu(x_{i-1})\bigg)\right].
                                                                \label{Gl13} 
\ee
This is the contribution to the rate of motion of the $i$-th step, due to
the difference between the chemical potentials of the $i$-th and the
$i - 1$-th step. In the same way the contribution of the difference
between the chemical potentials of the $i$-th and $i + 1$-th step to
the motion of the $i$-th step is
\be
- \frac{h_0}{k_{\mathrm{B}}T} \left( \frac{\partial h}{\partial x} \right)^{-1}\ 
 \left[\frac{\partial}{\partial x}\ \bigg(\Delta\mu(x_{i})\bigg)\right].
                                                                \label{Gl14}
\ee
Thus the total contribution of the variation of the chemical potential to
the rate of step motion is obtained by substituting Eqs.\ (\ref{Gl13}) and
(\ref{Gl14}) into Eqs. (\ref{Gl10a},\ref{Gl10b}) and (\ref{Gl7}),
\be
\frac{dx_i}{dt} = 
 - \frac{D_s \Omega h_0\ n^e_s\ \beta(x_i - x_{i-1})}
  {d_d(1+\beta)k_{\mathrm{B}}T}\ \frac{\partial}{\partial x}
   \left\{\left( \frac{\partial h}{\partial x}\right)^{-1}\ 
    \left[\frac{\partial}{\partial x}\bigg(\Delta\mu(x_i)\bigg)\right]\right\}.
                                                                \label{Gl15}
\ee
It can be seen from Eq.(\ref{Gl15}) that the rate of relaxation of a non-equilibrium
configuration of steps towards an ideal equidistant configuration (having a zero contribution
of the step-step repulsion to the chemical potential) is low when the parameter $\beta$
is small, i.e., when the ES barrier is high. This is easy to understand since the 
relaxation of the step configuration takes place by detachment of atoms from those
steps with high chemical potential, and their subsequent attachment to other steps with
low chemical potential. These atoms have to overcome the ES barrier either in the 
detachment, or in the attachment process.   

Substituting the expression (\ref{Gl15}) into the equation
\be
\frac{\partial h}{\partial t} = -\ h_0\ \frac{dx_i/dt}{x_i-x_{i-1}}
                                                                \label{Gl16}
\ee
used by Frank \cite{Frank62} in developing the kinematic theory of crystal
growth, one obtains
\be
\frac{\partial h}{\partial t} = \frac{\partial}{\partial x}
 \left\{ \sigma \ 
  \left[\frac{\partial}{\partial x}\bigg(\Delta\mu(x)\bigg)\right]\right\}
                                                                \label{Gl17}
\ee
where
\be
\sigma = \frac{D_s \Omega h^2_0\ n^e_s\ \beta}
       {d_d(1+\beta)k_{\mathrm{B}}T}
\left( \frac{\partial h}{\partial x} \right)^{-1}                                         \label{Gl18}
\ee
is a surface mobility relating the mass current in the curly brackets on the
right hand side of (\ref{Gl17}) to the gradient of the chemical potential
$\Delta \mu$. The proportionality of the mobility (\ref{Gl18})
to the inverse of the surface slope is intimately related \cite{Nozieres87,Liu96} 
to our assumption of a large kinetic length (slow detachment/attachment), 
i.e., case (b) [Eq.(\ref{Gl9b})]. Indeed, carrying out the same manipulations for the 
opposite case (a) [Eq.(\ref{Gl9a})], one arrives instead at the expression
\be
\label{mob2}
\sigma = \frac{D_s \Omega h_0 n^e_s \beta}{k_B T [\beta + (d_d/\lambda_s)^2]}
\end{equation}
which is independent of the surface slope.

Since the continuum limit of the expression (\ref{Gl5}) is \cite{Lancon90} 
\be
\Delta\mu = -\ \frac{6 \Omega A(T)}{h_0^2} \ \frac{\partial h}{\partial x}\ 
  \frac{\partial^2h}{\partial x^2},                              \label{Gl19}
\ee
Eq.~(\ref{Gl17}) can be presented as
\be
\frac{\partial h}{\partial t} + \frac{\partial}{\partial x}
 \left\{ C \left(\frac{\partial h}{\partial x} \right)^{-1}\ 
  \left[\frac{\partial}{\partial x}\left(\frac{\partial h}{\partial x}\ 
   \frac{\partial^2 h}{\partial x^2}\right)\right]\right\} = 0
                                                                \label{Gl20}
\ee
where
\be
C = \frac{6A(T)D_s \Omega^2\ n^e_s\ \beta}{d_d(1+\beta)k_{\mathrm{B}}T} =
\frac{3 l_0^3 \Omega D_s n_s^e \beta}{d_d(1 + \beta)}.
                                                                \label{Gl21}
\ee
A similar equation for the relaxation of the step bunches has been published 
previously \cite{Liu96}. It describes 
the relaxation of the fluctuations in the step density
due to the step-step repulsion. Equation (\ref{Gl20}) is highly
nonlinear and differs qualitatively from 
the linear evolution equation originally introduced by Mullins
\cite{Mullins59} to describe the relaxation of a crystal surface
above the roughening temperature.

\section{Continuum theory of bunch shapes}
\label{Continuum}

\subsection{The continuum evolution equation}
\label{Continuumequation}

It was shown in Ref.42 how the continuum limit for a set of step
equations of the linear form (\ref{Gl11}) can be carried out in an essentially
rigorous manner. The main idea is to perform the coarse graining operation
on the level of the linear step dynamics, where it can be done exactly, and
subsequently derive the evolution equation for the height profile $h(x,t)$ through
a nonlinear variable transformation. Combining the result of this procedure with
the continuum limit (\ref{Gl20}) for the relaxational dynamics, we obtain the
full evolution equation for our problem. It takes the form of a continuity equation,  
\be
\label{full}
\frac{\partial h}{\partial t} + \frac{\partial J}{\partial x} = 
- h_0 \hat R_e,
\ee
where the expression for the current is
\be
\label{current}
J = - \frac{\hat R_e h_0^2 (1 - \beta)}{2(1 + \beta)} \frac{1}{m} - \frac{\hat R_e h_0^3}{6}
\frac{1}{m^3} \frac{\partial m}{\partial x} + \frac{C}{2} \frac{1}{m} 
\frac{\partial^2}{\partial x^2} m^2.  
\ee
Here $m = \partial h/\partial x$ is the surface slope, which will be taken to be
positive. The first term on the right hand side of (\ref{current}) is the 
destabilizing, downhill current which is responsible for the step bunching instability,
while the last term describes the smoothening effect of the step-step repulsion,
as described above. The second term is the only one to break the reflection symmetry
($x \to - x$ and $m \to -m$) of the evolution equation. As will be shown later,
this term leads to different behavior at the upper and the lower edges of the bunch.
Following earlier work \cite{Krug97,Politi96}, we will refer to it as the symmetry-breaking
term. 

It is noteworthy that the continuum equation (\ref{full}) conserves the volume of the
film, apart from the constant sublimation rate on the right hand side, which does
not couple to the surface morphology. A dependence of the sublimation rate on 
the surface slope will be felt when the step spacing becomes comparable to the
diffusion length \cite{Burton51}. In this sense Eq.(\ref{full}) is valid
only to leading order in $l/\lambda_s$.

\subsection{Linear stability analysis}
\label{Linear}

It is straightforward to derive from (\ref{full},\ref{current}) the instability condition for a vicinal surface of slope $m_0 = h_0/l$. 
We set $h(x,t) = m_0 x - \hat R_e h_0 t 
+ \epsilon_q(x,t)$, where
$\epsilon_q(x,t) \sim e^{{\mathrm{i}} 
qx + \omega(q) t}$ is a perturbation of wavenumber
$q$, and expand (\ref{full},\ref{current}) to linear order in $\epsilon$. This yields
the expression
\be
\label{omegaq}
\omega(q) = \frac{\hat R_e h_0^2 (1 - \beta)}{2(1 + \beta) m_0^2} q^2
- C q^4 - \frac{\hat R_e h_0^3}{6 m_0^3} {\mathrm{i}} q^3
\ee
for the growth rate of the perturbation. The perturbation grows when
${\cal{R}}(\omega) > 0$, i.e. for wavenumbers $q < q_{\mathrm{max}} =
\sqrt{\hat R_e h_0^2 (1 - \beta)/[2(1 + \beta)m_0^2 C]}$, and perturbations
with wavenumber $q^\ast = q_{\mathrm{max}}/\sqrt{2}$ are maximally
amplified. The corresponding wavelength is
\be
\label{lambdaast}
\lambda^\ast = \frac{2 \pi}{q^\ast} = 4 \pi \sqrt{3 S} l,
\ee
where we have introduced the dimensionless quantity
\be
\label{S}
S = \frac{\beta}{1 - \beta} \; \frac{\lambda_s^2 \; l_0^3}{d_d \; l^4} = 
\frac{K_d K_u}{K_d - K_u} \frac{\tau}{l} \left( \frac{l_0}{l} \right)^{3}.
\ee
We will see below that the physical parameters of the problem enter
the properties of the bunch shape only in the combination (\ref{S}).
It is interesting to note that $S$ does not depend on the surface
diffusion constant $D_s$.   
The wavelength $\lambda^\ast$ determines the linear size of bunches
in the beginning of the bunching instability. Correspondingly the
number of steps in an incipient bunch is given by 
\be
\label{Nstar}
N^\ast = \lambda^\ast/l = 4 \pi \sqrt{3 S} \approx 21.8 \times \sqrt{S}. 
\ee
The imaginary part of the growth rate (\ref{omegaq}), which derives
from the symmetry-breaking term in (\ref{current}), does not affect the 
stability of the surface, but it induces a drift of fluctuations.

\subsection{The mechanical analog}
\label{Mechanics}

Our main goal in this section is to compute the shape of large, almost stationary
bunches from Eq.(\ref{current}).
For a stationary profile, the current (\ref{current}) is set to a constant $J_0$.
We neglect the symmetry-breaking term for now, and return to its relevance at the end
of the section. Introducing the quantity $u = m^2$, which is positive by construction,
the stationarity condition $J \equiv J_0$ 
can then be written in the familiar form
\be
\label{Newton}
\frac{C}{2} \frac{d^2 u}{dx^2} = B + J_0 u^{1/2} = - \frac{dV}{du}
\end{equation}
of a classical particle coordinate $u(x)$ moving in a potential $V(u) = - B u - (2/3) J_0 u^{3/2}$,
where $B = \hat R_e h_0^2 (1 - \beta)/[2(1 + \beta)] > 0$ and $J_0$ has to be chosen
negative. The bunch is a particle trajectory
which starts at $u = 0$ at ``time'' $x=0$, reaches a turning point $u = u_{\mathrm{max}}$
at time $x = L/2$,
and returns to $u = 0$ at time $x = L$. Here $L$ can be identified with the bunch
width,
and the minimal terrace length in the bunch introduced previously is given
by $l_{\mathrm{min}} = h_0/\sqrt{u_{\mathrm{max}}}$. 

\begin{figure}
\centerline{\epsfig{file=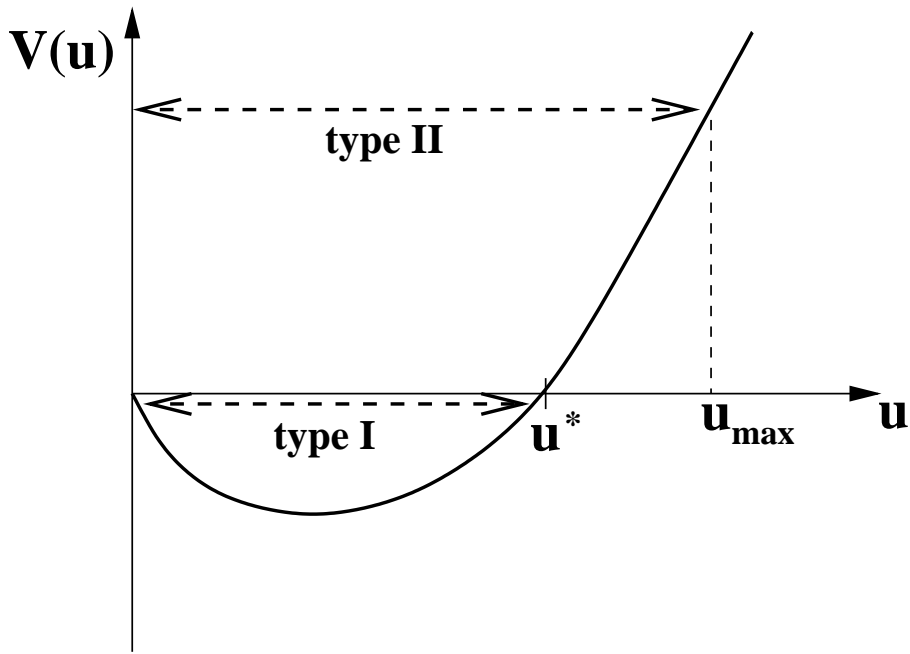,width=12.cm}}
\vspace*{0.5cm} 
\caption{\label{Fig1} Sketch of the potential $V(u)$ 
appearing in the mechanical analog. The dashed arrows illustrate
trajectories of type I and type II.}
\end{figure}

As can be seen in Fig. \ref{Fig1}, the potential $V(u)$ admits two types of periodic
trajectories. When the total particle ``energy'' ${\cal{E}} = (C/4)(du/dx)^2 + V(u)$
is negative, $u$ performs an essentially harmonic motion around the minimum of the potential.
This will be referred to as a \textit{type I} solution. 
The particle velocity $du/dx$ vanishes at both turning points, which translates into
the vanishing of the curvature $dm/dx$ of the surface profile. 
The height profile represented by such a trajectory 
is a periodic array of step bunches which is everywhere smooth. 
In contrast, particle
trajectories with positive total energy (\textit{type II} solutions) 
reach the reflecting ``hard wall'' at $u = 0$
at a finite speed. This implies that the surface curvature $dm/dx = (du/dx)/2m$ diverges
as $m \to 0$, such that the surface profile develops singularities of the type
\be
\label{PT}
h(x) - h(x_0) \sim (x - x_0)^{3/2}
\ee
near the bunch edges at $x = x_0$. This is the well-known Pokrovsky-Talapov law
which describes how the rounded parts of an equilibrium crystal shape join a flat 
facet \cite{Nozieres91}, and is reflected in the scaling of the terrace sizes
at the edges of the bunch (see Sect.\ref{Scaling}). Because of the singularities
at the bunch edges, type II height profiles have finite support and cannot be continued
to describe a periodic array of bunches.

We now show that the two types of trajectories imply different scaling properties
for the step bunches.  
For a trajectory of type I to satisfy the boundary
conditions $u(0) = u(L) = 0$ for $L \to \infty$, 
it is necessary to set the particle energy
${\cal{E}} = 0$. The right turning point is then located at 
the value $u^\ast = (3B/2J_0)^2$ at which $V(u^\ast) = 0$, and the period of
the trajectory is readily seen to be of the order of $L \sim \sqrt{BC}/\vert J_0 \vert$.
Bunches of arbitrary (large) lateral size $L$ can therefore be accommodated
only by treating $J_0$ as a free integration constant, and to let $J_0 \sim 1/L
\to 0$ for $L \to \infty$. This is the procedure adopted in Ref.34
for a slightly different case. In the present context it implies the
scaling relation $l_{\mathrm{min}} \sim 1/\sqrt{u^\ast} \sim \vert J_0 \vert \sim 1/L$,
and since the number of steps in the bunch is of the order $N \sim L/l_{\mathrm{min}}$,
we find that $l_{\mathrm{min}} \sim N^{-1/2}$. 
For future reference we record the full expressions for $L$ and $l_{\mathrm{min}}$,
which read 
\be
\label{LtypeI}
L = (216 \, S)^{1/4} \; N^{1/2} \; l ,\;\;\;\;
l_{\mathrm{min}} = \frac{2}{3} \frac{L}{N} = \left( \frac{128 S}{3} \right)^{1/4} \frac{l}{N^{1/2}}.
\ee

For type II trajectories with ${\cal{E}} > 0$, it is possible to make the period
arbitrarily large while keeping $J_0$ fixed, simply by increasing the energy.
For large bunches we then have $u_{\mathrm{max}} \gg u^\ast$, and the decreasing
part of the potential becomes irrelevant. The equation for the profile reduces
to the form  
\be
\label{Nozieres} 
C \frac{d}{dx} \left( m \frac{dm}{dx} \right) = - \vert J_0 \vert m,
\ee
which was first studied by Nozi\'eres \cite{Nozieres87}, and leads
to the scaling law\cite{Liu98a,Stoyanov98a,Nozieres87} 
$l_{\mathrm{min}} \sim N^{-2/3}$ 
(see Sect.\ref{Scaling} for a detailed
derivation). Anticipating the results of the numerical solution of the discrete
step equations in Sect.\ref{Numerics}, it turns out that this scaling
law is in agreement with the numerical data for $l_{\mathrm{min}}$, 
while the prediction (\ref{LtypeI}) for type I trajectories is not. We conclude, therefore,
that the type II trajectories of (\ref{Newton}), with singular behavior
at $u = 0$, are the relevant solutions for the description of step bunches. 
This immediately
raises the question of how the current $J_0$, which then no longer can be treated as
an integration constant, should be determined. The answer will be given in the next subsection.

We add a final remark comparing the two types of solutions. 
On purely mathematical grounds, at first sight the type I solutions may seem to be preferable,
because they avoid the singularities at the bunch edges and allow to describe a periodic
array of many bunches. This is in fact not the case. 
For a periodic type I profile, the end of one bunch (the point where $m=u=0$) defines the beginning
of the next. As, according to (\ref{LtypeI}), the bunches steepen with
increasing size, the mean slope of the surface also increases without bound. 
This is in contradiction to the time evolution of a real surface, for which the mean
slope is fixed and the steepening of the bunches is compensated by the growth of large
flat regions between the bunches (see Fig.\ref{Fig0}). For type I solutions, the region
between bunches shrinks to a point. Therefore they cannot be taken at face value as a 
global description of a surface with many bunches. Just like for the type II solutions, 
which terminate in singularities, also type I solutions have to be complemented by
a separate description for the flat regions between bunches.

\subsection{The mean surface current}        
\label{Current} 

Our strategy will be to fix $J_0$ by analogy with two related problems, which are mathematically
equivalent to the present one (see Sect.\ref{EquivalentProblems} for further discussion). 
The two problems are the linear step growth model considered
in Ref.42, and the model of surface electromigration in the attachment/detachment-limited regime 
considered in Ref.24. Indeed, the step equations derived by Liu and Weeks \cite{Liu98a}
for surface electromigration in the presence of desorption but without Ehrlich-Schwoebel barriers
are (apart from the different physical meaning of the coefficients) identical to our equations 
(\ref{Gl11}). By appealing to the analogy with surface electromigration, we can associate
a microscopic current $j_i$ with the terrace between the steps $i$ and $i+1$, which is
given by the expression
\be
\label{microcurrent}
j_i = - h_0 \hat D [\Delta \mu(x_{i+1}) - \Delta \mu(x_i)] - \frac{(1 - \beta)}{2(1 + \beta)}
h_0 \hat R_e (x_{i+1} - x_i).
\ee
Here the abbreviation $\hat D = D_s \Omega n_s^e \beta/[d_d k_{\mathrm{B}}T (1 + \beta)]$ has been
introduced. The physical meaning of this current is clear in the context of surface electromigration --
it is simply the current of adatoms driven across the terrace by the combined action 
of the electromigration force and the chemical potential gradient. The notion of a surface current
induced by attachment asymmetry is also well established in the context of epitaxial
growth \cite{Michely03,Krug97,Villain91,Krug93,Politi00}. 
It is less evident that the concept can be extended
to sublimation, where the atoms detached from the steps do not necessarily remain on the terrace.
However, as was noted above in Sect.\ref{Continuumequation}, we are working here in a limit
where the diffusion length $\lambda_s$ is very large, and hence the mass transport is essentially
confined to the surface.
Mathematically, the derivation in Ref.42 shows that for a general set of step equations 
which are linear in the terrace widths,
\be
\label{linear}
\frac{d x_i}{dt} = \gamma_u (x_{i+1} - x_i) + \gamma_d (x_i - x_{i-1}),
\ee
the leading order (stabilizing or destabilizing)
part of the surface current is proportional to the asymmetric combination $\gamma_u - \gamma_d$
of the contributions from the two terraces, while the symmetric combination 
$\gamma_u + \gamma_d$ gives rise to the overall growth (or sublimation) rate of the surface,
and to the symmetry-breaking part of the surface current.  

For a perfect step train with constant step spacing $l$, the current (\ref{microcurrent}) is equal
to 
\be
\label{Jflat}
J_{\mathrm{flat}} = - \frac{1 - \beta}{2(1 + \beta)} h_0 \hat R_e l,
\ee 
which is just the first term in (\ref{current}) evaluated at slope $m = h_0/l$. Following Liu and
Weeks\cite{Liu98a}, we now argue that the expression (\ref{Jflat}) remains valid also for a periodic
array of step bunches. Each bunch contains $N$ steps, and hence consists of 
$N-1$ short terraces. The width of each bunch is $L$, and the
bunches are separated by terraces of length $L_t$. We denote by $\Delta \mu_-$ and 
$\Delta \mu_+$ the values of the chemical potential at the lower and upper edges of the bunch.
Summing the expression (\ref{microcurrent}) across the bunch, we find that the current
in the bunch is equal to 
\be
\label{bunchcurrent}
j_b = - \frac{h_0}{N-1} \left[ \hat D (\Delta \mu_+ - \Delta \mu_-) + 
\frac{(1 - \beta)}{2(1 + \beta)} \hat R_e L \right],
\ee  
while the current on the terrace is
\be
\label{terracecurrent}
j_t = - h_0 \hat D (\Delta \mu_- - \Delta \mu_+)- \frac{(1- \beta)}{2(1 + \beta)} h_0 \hat R_e L_t.
\ee
Stationarity requires that $j_b = j_t$, which yields an expression for the chemical potential
difference $\Delta \mu_+ - \Delta \mu_-$. Inserting this back into (\ref{bunchcurrent}) or
(\ref{terracecurrent}), we find the simple result
\be
\label{bothcurrent}
j_b = j_t = - \frac{(1 - \beta)}{2(1 + \beta)} h_0 \hat R_e \frac{L + L_t}{N},
\ee
which coincides with (\ref{Jflat}). Thus the overall current remains constant, at its
value for a regular step train, throughout the bunching process, and the appropriate
expression to use for $J_0$ in (\ref{Newton}) or (\ref{Nozieres}) is $J_{\mathrm{flat}}$. 

It is possible to argue for this choice of $J_0$ also without reference to the
discrete step dynamics. Indeed, setting $J_0 = J_{\mathrm{flat}}$ ensures that
the minimum of the potential $V(u)$ is located at the value $u = (h_0/l)^2$ corresponding
to the mean surface slope. In this way the initial regular step train, which is
clearly an (unstable) stationary solution of the discrete step dynamics, is retained
as a solution also in the stationary continuum equation.   

\subsection{Derivation of the scaling laws}
\label{Scaling}

The solution of the mechanical problem (\ref{Newton}) allows to express the
bunch width $L$ and the bunch height $H$ in terms of the maximal slope
$m_{\mathrm{max}} = \sqrt{u_{\mathrm{max}}}$. 
Using energy conservation and neglecting the linear term
$-B u$ in the potential $V(u)$, we obtain the expressions
\be
\label{bunchL}
L = \sqrt{\frac{3 C}{2 \vert J_0 \vert}} \int_0^{u_{\mathrm{max}}} \frac{du}{\sqrt{
u_{\mathrm{max}}^{3/2} -  
u^{3/2}}} \approx 2.11 \sqrt{\frac{C}{\vert J_0 \vert}} m_{\mathrm{max}}^{1/2}
\ee
and 
\be
\label{bunchH}
H = \sqrt{\frac{3C}{2 \vert J_0 \vert}} \int_0^{u_{\mathrm{max}}} 
\frac{\sqrt{u}\; du}{\sqrt{u_{\mathrm{max}}^{3/2} - u^{3/2}}} = 
\sqrt{\frac{8}{3}} \sqrt{\frac{C}{\vert J_0 \vert}} m_{\mathrm{max}}^{3/2}.
\ee
The dimensionless coefficient in (\ref{bunchL}) was obtained by numerically
evaluating the corresponding integral.
Inserting the expressions (\ref{Gl21}) and (\ref{Jflat}) for $C$ and 
$J_0$, respectively, we can write  
the minimal terrace length $l_{\mathrm{min}}$
and the bunch width $L$ in terms of the number of steps and dimensionless
ratios of length scales, as 
\be
\label{lminN}
\frac{l_{\mathrm{min}}}{l} = 2^{4/3} \left(\frac{\beta}{1 - \beta} \right)^{1/3}
\left( \frac{l}{d_d} \right)^{1/3} \left( \frac{\lambda_s}{l} \right)^{2/3}
\left(\frac{l_0}{l} \right) N^{-2/3} = 2^{4/3} S^{1/3} N^{-2/3}
\ee
and
\be
\label{LN}
\frac{L}{l} \approx 3.25 \left( \frac{\beta}{1 - \beta} \right)^{1/3} 
\left( \frac{l}{d_d} \right)^{1/3} \left( \frac{\lambda_s}{l} \right)^{2/3}
\left( \frac{l_0}{l} \right) N^{1/3} = 3.25 \; S^{1/3} N^{1/3}.
\ee
Equations (\ref{lminN}) and (\ref{LN}) are the central results of this section, and will
be compared to numerical simulations of the discrete step model in Sect.~\ref{Numerics}.
Together (\ref{lminN}) and (\ref{LN}) imply the universal relation
$L/l_{\mathrm{min}} \approx 1.29 N$, independent of all physical parameters. This shows
that the minimal terrace length is only a factor 0.78 smaller than the mean terrace
length $L/N$ within the bunch, hence most terraces in the bunch have a size of order
$l_{\mathrm{min}}$. 

A separate scaling law holds for the size of the first (and last) terrace at the edges
of the bunch. To derive it, we need to determine the precise bunch profile near 
$u=0$, which is of the general form (\ref{PT}). Since the mechanical potential $V(u)$
vanishes at $u=0$, energy conservation implies that $(C/4)(du/dx)^2 = {\cal{E}} = 
V(u_{\mathrm{max}})$ for $u \to 0$. Integrating this from the bunch edge at $x=0$ yields
the slope profile $m(x) = (4 {\cal{E}}/C)^{1/4} x^{1/2}$, and correspondingly
the height profile $h(x) = (2/3)(4 {\cal{E}}/C)^{1/4} x^{3/2}$. 
There is an ambiguity in how to estimate the size $l_1$ of the first terrace --
a quantity manifestly related to the discreteness of the surface in the vertical direction --
from these continuous profiles. Natural choices would be to require that 
(i) $h(x = l_1) = h_0$, (ii) $m(x = l_1) = h_0/l_1$ or (iii) $m(h = h_0) = h_0/l_1$.
It is easy to check that all three choices imply a scaling relation
of the form $l_1 \sim S^{1/3} N^{-1/3}$, but with
different numerical coefficients. Because the continuum equation is derived \cite{Krug97} 
from a set of discrete equations for the terrace sizes (the inverse slopes) as a function
of the layer height, we claim (and confirm numerically in Sect.\ref{Numerics})
that choice (iii) is the appropriate one. This results in       
\be
\label{l1}
\frac{l_1}{l} = 4^{1/3} \left( \frac{\beta}{1-\beta} \right)^{1/3}
\left( \frac{\lambda_s}{l} \right)^{2/3} \left( \frac{l}{d_d} \right)^{1/3} 
\left( \frac{l_0}{l} \right) N^{-1/3} = 4^{1/3} \; S^{1/3} \; N^{-1/3}.
\ee
The same kind of analysis can be carried out for the
type I solutions discussed in Sect.\ref{Mechanics}. One finds that 
$l_1/l \sim S^{1/4}$ independent of the bunch size $N$.   

\subsection{General step interaction}

The above considerations are easily generalized to different values for the exponent $n$ describing
the decay of the step-step interaction as $1/l^n$. 
The expression (\ref{Gl5}) for the chemical potential at
step $i$ then becomes 
\be
\label{chempotgen}
\frac{\Delta\mu(x_i)}{k_{\mathrm{B}}T} = -\ \left(\frac{l_0}{x_{i+1}-x_{i}}\right)^{n+1} +
        \left(\frac{l_0}{x_i - x_{i-1}}\right)^{n+1}
\ee
with $l_0 = (n \Omega A/k_{\mathrm{B}} T)^{1/(n+1)}$, and going through the 
manipulations of Sect.\ref{Contrelax} one obtains the generalized relaxation equation 
\be
\frac{\partial h}{\partial t} + \frac{\partial}{\partial x}
 \left\{ C_n \left(\frac{\partial h}{\partial x} \right)^{-1}\ 
  \left[\frac{\partial}{\partial x}\left(\frac{\partial h}{\partial x}\ \right)^{n-1} 
   \frac{\partial^2 h}{\partial x^2}\right]\right\} = 0
                                                                \label{Relaxgen}
\ee
with 
\be
\label{Cgen}
C_n = \frac{(n+1) l_0^{n+1} \Omega D_s n_s^e \beta h_0^{2-n}}{d_d (1 + \beta)}.
\ee
The analysis of the resulting full continuum equation leads to the scaling relations \cite{Sato99}
\be
\label{genscaling}
l_{\mathrm{min}} \sim N^{-2/(n+1)}, \;\;\; 
L \sim N^{(n-1)/(n+1)}, \;\;\;
l_{1} \sim N^{-1/(n+1)}
\ee
for type II solutions. 
Specifically, the relations (\ref{lminN},\ref{l1}) generalize to 
\be
\label{lminNgen}
l_{\mathrm{min}}/l  = (16 S_n)^{1/(n+1)} N^{-2/(n+1)} = (16 S_n/N^2)^{1/(n+1)} , \;\; 
l_1/l = (4S_n/ N)^{1/(n+1)},
\ee  
where the dimensionless parameter $S_n$ is defined by
\be
\label{Sn}
S_n = \frac{\beta}{1 - \beta} \frac{\lambda_s^2 l_0^{n+1}}{d l^{n+2}} =
\frac{K_d K_u}{K_d - K_u} \frac{\tau}{l} \left( \frac{l_0}{l} \right)^{n+1}.
\ee
Comparing the two expressions in (\ref{lminNgen}) it is seen that
$l_{\mathrm{min}}$ becomes smaller than $l_1$ only for $N > 4$; this 
can be viewed as a necessary condition for the onset of scaling.
If we require in addition that $l_1/l < 1$, it follows that such small bunches can form
only provided $S_n < 1$.   

For type I solutions the scaling relations corresponding to (\ref{genscaling})
read
\be
\label{ScalinggenI}
L \sim S_n^{1/(n+2)} N^{n/(n+2)}, \;\;\; l_{\mathrm{min}} \sim S_n^{1/(n+2)} N^{-2/(n+2)}.
\ee

\subsection{Corrections to the asymptotic behavior}    
\label{Corrections}

The scaling laws derived in the preceding subsection are valid asymptotically for large bunches,
when the two approximations made in their derivation are well justified: First,
neglecting the symmetry-breaking contribution to the current (\ref{current}), and second,
neglecting the constant term in the particle equation (\ref{Newton}). The second
approximation is valid provided $u_{\mathrm{max}} \gg u^\ast$, where $u^\ast = (3 B/2J_0)^2$ 
is the value of $u$ at which $V(u^\ast) = 0$. Inserting the expression (\ref{Jflat})
for $J_0$, we see that the corresponding slope $m^\ast = \sqrt{u^\ast}$ is simply
of the order of the mean slope $h_0/l$ of the surface. Thus the linear term in $V(u)$
can be ignored throughout the bunch. 

Including the symmetry-breaking term, the stationarity condition $J = J_0$ 
can be brought into the form
\be
\label{Newton2} 
S \frac{d^2 v}{dy^2} = \frac{1}{3} (1 - \sqrt{v}) + \frac{1 + \beta}{18(1 - \beta)}
\frac{1}{v^{3/2}} \frac{dv}{dy},
\ee
where $v = (l/h_0)^2 u$ is a dimensionless version of $u$, normalized such that
the mean slope corresponds to $v = 1$, and $y = x/l$ is the dimensionless spatial coordinate. 
In the mechanical analog of Sect.\ref{Mechanics}, the symmetry-breaking term corresponds
to a friction force which, because of its sign, acts ``backward'' in time. 
The friction force is proportional to $v^{-3/2}$. Hence it is
negligible deep inside the bunch, where $v \gg 1$, but becomes important near the edges
of the bunch. 

\begin{figure}
\centerline{\epsfig{file=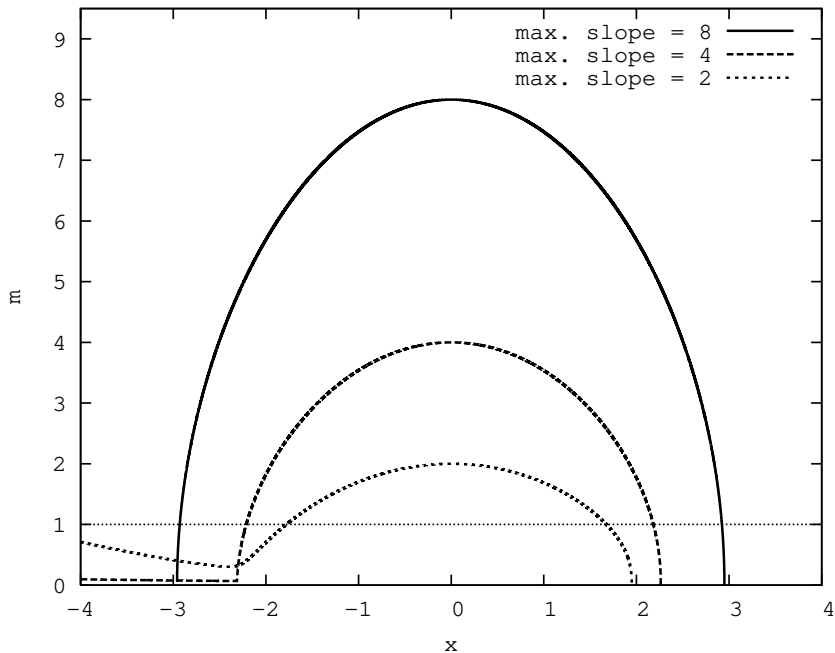,width=9.cm,angle=-90}}
\vspace*{0.5cm} 
\caption{\label{Fig2} Stationary bunch shapes computed 
by numerical integration of Eq.(\ref{Newton2}) with $S = 0.14$ and 
$\beta = 0.01$. The figure shows the dimensionless slope $(l/h_0) m(x)$
as a function of the rescaled coordinate $x/l$ for different values
of the maximum slope. The horizontal dotted line shows the value
$m = h_0/l$ corresponding to the mean slope of the unperturbed surface.}
\end{figure}

It is instructive to follow the solution of Eq.(\ref{Newton2}) starting from 
the center of the bunch, where $v(y)$ takes its maximum value $v_{\mathrm{max}}$ and 
$dv/dy = 0$. Integrating forward in ``time'' $y$, towards the upper edge
of the bunch, the friction force adds to the acceleration
of the particle towards $v=0$, which is therefore reached earlier than in the absence
of the symmetry-breaking term. Moreover the singular behavior at $v=0$ is altered:
Balancing the symmetry-breaking term against the intertial term on the left hand side
of (\ref{Newton2}), it is straightforward to show that the standard behavior (\ref{PT}) of the
height profile is modified into 
\be
\label{PT2}    
h(x_0) - h(x) \sim (x_0 - x)^{4/3}.
\ee
Conversely, when moving backward in time (towards the lower edge of the bunch)
the particle is delayed by the friction force. Since the friction coefficient
diverges as $v \to 0$, the point $v=0$ is never reached. Instead, the trajectory
bounces back and approaches the stable potential minimum at $v = 1$ in an overdamped
or damped oscillatory manner. Since the parameter $S$ plays the role of the particle
mass in (\ref{Newton2}), the effects of friction increase with decreasing $S$, while
they decrease when increasing the initial particle energy (i.e., the value of 
$v_{\mathrm{max}}$, and, hence, the size of the bunch).  

Thus the symmetry-breaking term modifies the nature
of the solutions of (\ref{Newton2}) in a qualitative way: Whereas the      
relevant solutions of the frictionless particle problem (\ref{Newton}) have finite
support in $x$ (the trajectory returns to $u=0$ in a finite time), the solutions
of (\ref{Newton2}) extend all the way to $y = - \infty$, where $v$ attains the limiting
value $v = 1$. In physical terms this implies that the bunch width $L$ can no longer be 
sharply defined within the continuum theory. Nevertheless, the numerical solutions
of (\ref{Newton2}) depicted in Fig. \ref{Fig2} show that this effect is completely 
negligible already for moderately large bunches and physically relevant values of 
$S$. Deviations from the solution of the frictionless equation (\ref{Newton}) occur only
in the range $v \ll 1$, which is irrelevant for the description of actual bunches
(recall that $v=1$ corresponds to the mean slope $h_0/l$ of the unperturbed surface). 
This conclusion is supported by a scaling analysis in the spirit of Ref.36 
(see Sect.\ref{Universality}).

We will see in Sect.\ref{Numerics} that the left-right symmetry of the bunches is
indeed broken in a way that is qualitatively reminiscent of the solutions of (\ref{Newton2}) with
very small $S$. However, the sign of the observed symmetry-breaking is opposite to that predicted by 
(\ref{Newton2}), and we will argue that its origin is in fact completely different.     

\section{Equivalent Problems}
\label{EquivalentProblems}

We noted already in Sect.\ref{Current} that the equations of step motion 
(\ref{Gl7},\ref{Gl10a},\ref{Gl10b}) are 
mathematically equivalent to appropriate limiting cases of those obtained
for step bunching instabilities induced by electromigration and growth in the
presence of inverse Ehrlich-Schwoebel barriers. Here we elaborate on that 
observation and translate the results derived from the continuum theory 
to the different physical contexts.

\subsection{Electromigration}
\label{Electromigration}

Discrete and continuum equations for electromigration-induced step bunching 
in the attachment/detachment limited regime have been 
derived by Liu and Weeks \cite{Liu98a}, and their equations are readily seen to be 
of the same form as ours. In our setting, an electromigration force $F$ acting on the
adatoms can simply be added to the chemical potential gradient on the right hand side
of (\ref{Gl17}). 
This gives rise to an additional contribution $J_F = \sigma F$ to the surface current 
$J$ in (\ref{full}), which is also inversely proportional to the slope $m$ 
[see the expression (\ref{Gl18}) for $\sigma$ in the attachment/detachment limited case],
and which is destabilizing for $F < 0$ (force in the downhill direction). 
In the absence of an Ehrlich-Schwoebel effect ($\beta = 1$), the electromigration
current is the only destabilizing contribution. The results of Sect.\ref{Continuum} carry
directly over to this case, once the dimensionless parameter $S$ has been identified
along the lines of Sect.\ref{Linear}. One finds the simple result
\be
\label{Semig} 
S = \frac{\Omega A}{F l^4}
\ee
and hence from (\ref{lminN},\ref{LN},\ref{l1}) we obtain the predictions
\be
\label{Lengthsemig}
l_{\mathrm{min}} = \left( \frac{16 \Omega A}{F l} \right)^{1/3} N^{-2/3},\;\;\;
L \approx 3.25 \left( \frac{\Omega A}{F l} \right)^{1/3} N^{1/3}, \;\;\;\;
l_1 = \left( \frac{4 \Omega A}{F l} \right)^{1/3} N^{-1/3}.
\ee
Similar formulae have been reported previously in the literature.
Sato and Uwaha derived the results \cite{Sato99}
\be
\label{Sato}
L/N \approx 2.59 \; \left( \frac{\Omega A}{F d} \right)^{1/3} N^{-2/3}, \;\;\;
l_{\mathrm{min}} = \left( \frac{8 \Omega A}{F d} \right)^{1/3} N^{-2/3}, \;\;\;
l_1 = \left( \frac{2 \Omega A}{F d} \right)^{1/3} N^{-1/3},
\ee
which are of the same form as the expressions in (\ref{Lengthsemig}), with the kinetic
length $d = d_u = d_d$ replacing the mean step spacing $l$. This reflects the fact
that Sato and Uwaha work in the diffusion-limited regime $d < l$.
  
Stoyanov and Tonchev \cite{Stoyanov98a} have developed a continuum description for
electromigration-induced step bunching in the diffusion-limited regime.
Assuming the relation $d = a_\parallel$ for the kinetic length, which
holds for non-permeable steps in the absence of an additional barrier
against attachment, they obtained the evolution equation 
\be
\label{rho=1equation}
\frac{\partial h}{\partial t} + \frac{\partial}{\partial x} 
\left[\tilde B m + \frac{\tilde C}{2} \frac{\partial^2}{\partial x^2} m^2 \right]
=0,
\ee
where $m = \partial h/\partial x$ is the surface slope, and the coefficients
$\tilde B$ and $\tilde C$ are given by
\be
\label{BCtilde}
\tilde B = - \frac{2 D_s n_s^e F \Omega a_\perp}{k_{\mathrm{B}} T}, \;\;\;\;
\tilde C = \frac{6 D_s n_s^e \Omega^2 A}{h_0 k_{\mathrm{B}} T}.
\ee
The surface is unstable when the force is in the downhill direction, 
$F < 0$ and $\tilde B > 0$. 
The relaxation term in (\ref{rho=1equation}) is simply the product of
the chemical potential variation (\ref{Gl19}) and the expression (\ref{mob2}) for the
mobility in the diffusion-limited case, where it has been used that
$d_d \ll \lambda_s$ and $\beta = 1$.

The analysis of stationary solutions of (\ref{rho=1equation}) proceeds along the 
lines of Sect.\ref{Mechanics}. In fact, the mechanical analog resulting
from setting the square bracket in (\ref{rho=1equation}) equal to a constant
current $\tilde J_0$ is formally
identical to (\ref{Newton}), with 
the potential $V(u) = - \tilde J_0 u + \tilde B u^{3/2}$. Despite the
formal similarity, however, the problem differs from that considered in
Sect.\ref{Mechanics} in one important respect:
Since the overall surface current is downhill, we have to chose $\tilde J_0 < 0$,
and hence both terms in the potential are positive; the potential has no minimum, and
type I trajectories do not exist. For the type II trajectories
the term proportional to $\tilde J_0$ becomes irrelevant at large slopes, and hence
one may as well set $\tilde J_0 = 0$, as was done in Ref.29. 
The bunch shape is then given by the solution of (\ref{Nozieres}), and the results
of Sect.\ref{Scaling} can be taken over. This yields the scaling relations
\be
\label{Stoyan}
L/N \approx 2.58 \left( \frac{a_\parallel A}{F} \right)^{1/3} N^{-2/3}, \;\;\;\;
l_{\mathrm{min}} = 2 \left( \frac{a_\parallel A}{F} \right)^{1/3} N^{-2/3}.
\ee
Apart from a small difference in the dimensionless prefactor, the expression 
for $L/N$ is identical to the one reported in Ref.29. 

Relations of the form (\ref{Stoyan}) have 
been used in the interpretation of several experiments on silicon surfaces,
where a scaling of the minimal terrace size \cite{Fujita99}
and the mean terrace size \cite{Homma00} as $N^{-2/3}$ was observed. 
The similarity between the expressions (\ref{Lengthsemig},\ref{Sato},\ref{Stoyan}) implies
that it is not possible to distinguish between attachment/detachment limited kinetics
and diffusion-limited kinetics on the basis of the observed scaling; see
Sect.\ref{Universality} for further elucidation of this point. However, the resulting
estimate for the ratio $A/F$ depends crucially on which kinetic regime is assumed. 
Consider for example the results obtained by Fujita, Ichikawa and Stoyanov \cite{Fujita99} 
for the maximum bunch slope $h_0/l_{\mathrm{min}}$ at
1250$^{\mathrm{o}}$ C. Using the relation (\ref{Stoyan}) for the diffusion-limited
regime yields the estimate $F/A \approx 3 \times 10^{-6} \mathrm{nm}^{-2}$; this is somewhat
larger than the value reported in Ref.31, because in that work the authors
used an expression for the mean terrace width $\bar l = L/N$.
Because of the additional factor of $l$, application of (\ref{Lengthsemig}) 
yields instead $F/A \approx 6 \times 10^{-8} \mathrm{nm}^{-2}$, which is smaller by
a factor of 50.
Correspondingly the effective valence $Z^\ast$ of the silicon adatoms, defined
through the relation $F = Z^\ast e E$ between the electromigration force and the electric
field $E$, will be smaller by the same factor. 

\subsection{Growth with inverse Ehrlich-Schwoebel barriers}

Growth in the presence of an inverse Ehrlich-Schwoebel effect is described by the 
stationary diffusion equation (\ref{Gl2}) with $R > 0$ and
$1/\tau_s = 0$, and the boundary conditions (\ref{Gl3})
with $K_u > K_d$, i.e $\beta > 1$. While the possibility of inverse ES
barriers is debatable on the microscopic level, the inverse ES effect is
may serve as a useful effective description of more complex step bunching
mechanisms \cite{Myslivecek02}.   
The equations of step motion can be found, e.g., in Ref.12. We consider 
the limiting case of fast attachment to the descending step and slow attachment
to the ascending step, i.e. $d_d \gg x_i - x_{i-1} \gg d_u$. In this limit only
the upper terrace contributes to the growth
of the step, and the destabilizing part of the dynamics reduces to the linear form (\ref{linear}) 
with $\gamma_d = 0$ and $\gamma_u = - R \Omega$ 
(note that in our setup the steps recede during growth and the upper terrace trails
the step). The continuum equation is thus of the same form as in sublimation
and electromigration, and the results of Sect.\ref{Continuum} carry over with the identification
\be
\label{Sgrowth}
S = \frac{D_s n_s^e l_0^3}{R d_d l^4}
\ee
of the dimensionless parameter. The application of the continuum theory to this problem
will be the subject of a separate publication \cite{Tonchev03}.        

\section{Numerical analysis of the discrete step dynamics}
\label{Numerics}

\begin{figure}
\centerline{\epsfig{file=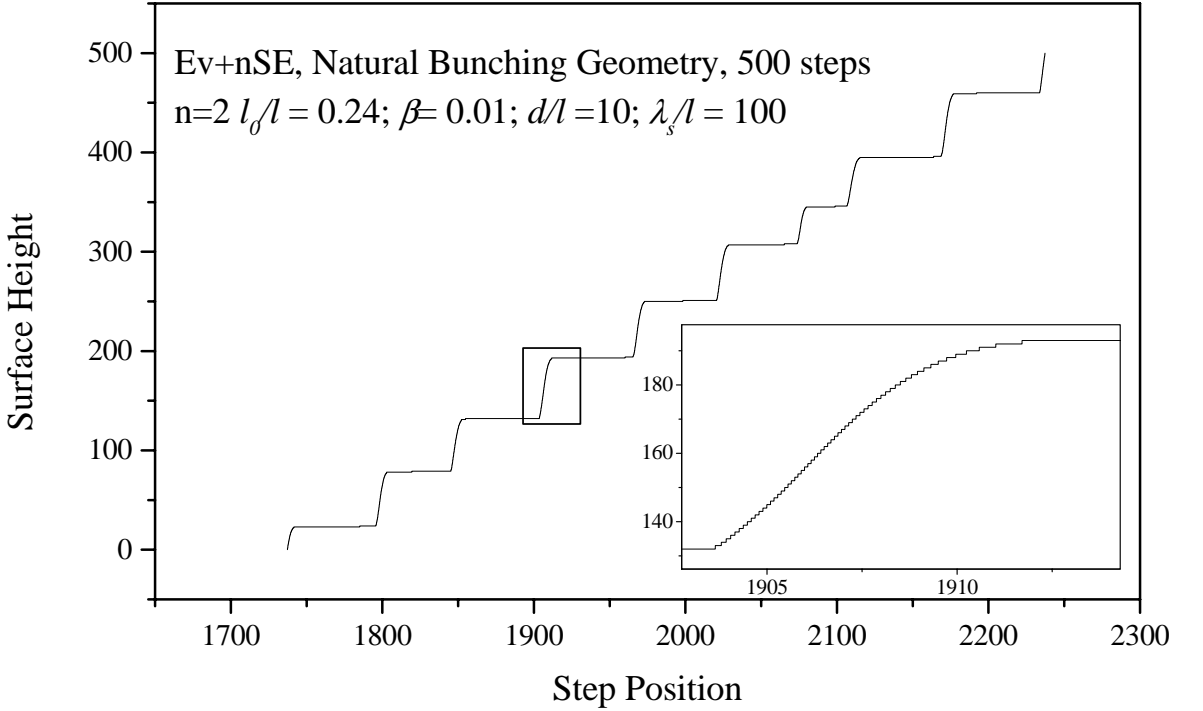,width=14.cm,angle=-90}}
\caption{\label{1} Profile of a crystal surface after some time of
sublimation. The inset shows the enlargement of an individual bunch.} 
\end{figure}

Extensive numerical simulations of the discrete step dynamics have been carried out
to test the predictions of the continuum theory. In this section we report
on simulations for the sublimation problem; a comprehensive numerical study
of growth in the presence of an inverse Ehrlich-Schwoebel effect will be
presented elsewhere \cite{Tonchev03}.  
We work with cyclic boundary conditions,
$x_{M+1} = x_1 + Ml$ where $l$ is the average interstep distance of
the vicinal surface and $M$ is the total number of steps, and prepare
the system in one of two kinds of initial conditions.
Under \textit{natural bunching conditions}, 
the integration is started 
from a vicinal surface with steps which
slightly deviate from their regular positions. This leads to a
surface consisting of many bunches of steps separated by large terraces,
which slowly coarsens 
(Fig.\ref{1}). 
On the other hand a \textit{single bunch}
can be prepared by chosing an initial step
configuration, corresponding to a bunch, which contains almost all
of the steps in the system. The integration then provides results for
the steady state shape of the bunch and the average value of the
number $N$ of steps in it (the remaining $M - N$ steps are
single -- they are crossing the large terrace between the front
edge of the bunch and its tail in our single bunch setup with cyclic boundary
conditions). In both cases we assume that a
given step belongs to the bunch when the distance to at least one
of the neighbouring steps is smaller than 0.75 $l$. This definition
is, in fact, arbitrary and it introduces some ambiguity in the results.
The dependence of the minimum distance $l_{\min}$ between the steps in the bunch
on the number $N$ of steps is less affected by this ambiguity than that of the total
bunch width $L$, as in the former case only $N$
is influenced by the definition, whereas $l_{\mathrm{min}}$ is precisely
determined. 

\begin{figure}
\centerline{\epsfig{file=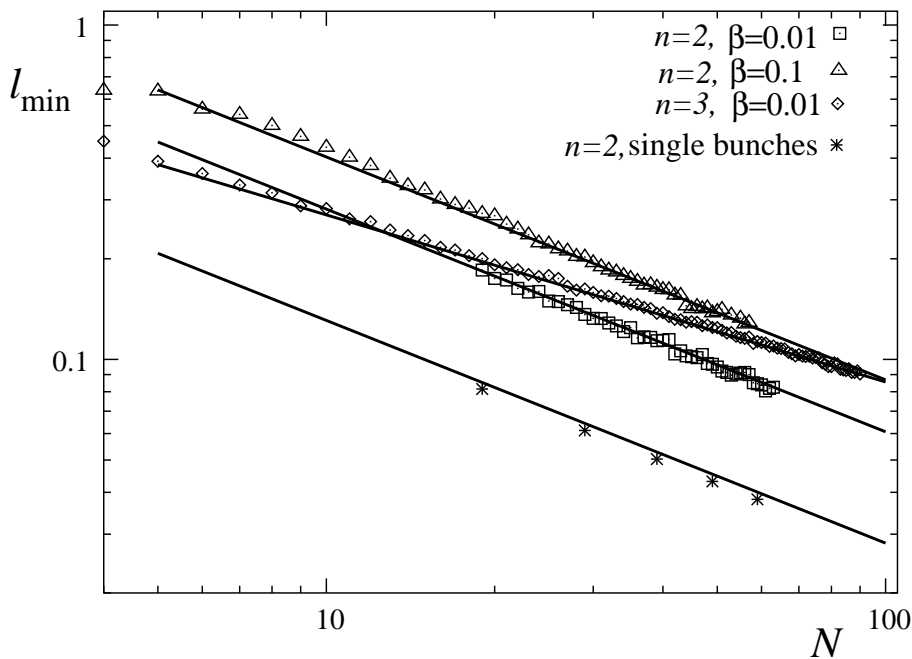,width=12.cm}}
\vspace*{0.5cm}
\caption{\label{2} 
Numerical data for the minimum interstep distance, measured in units
of $l$, as a function of bunch size. Open symbols show
results obtained in the natural bunching geometry with 500 steps, while asterisks
show data obtained from computations with a single bunch.
The interaction strength was
$l_0/l = 0.24$ in all cases. Open squares and diamonds show data for $d_d/l = 10$
and $\lambda_s/l = 100$, triangles show data for $d_d/l = 150$ and
$\lambda_s/l = 200$, and asterisks show data for $d_d/l = \lambda_s/l = 100$;
other parameters are given in the figure. Bold lines show the 
theoretical prediction (\ref{lminNgen}) for $l_{\mathrm{min}}$.}
\end{figure}

We first present results from the 
numerical integration of the sublimation problem
in case (b) of Sect.\ref{Equationsofmotion}, i.e. (\ref{Gl7}) was used
with $v_u$ and $v_d$ given by (\ref{Gl10a},\ref{Gl10b}), with the destabilizing
terms depending linearly on the terrace widths.
The equations of
step motion contain 4 dimensionless parameters:
$\lambda_s/l,\; d_d/l, \; l_0/l$ and
$\beta = K_u/K_d$. It is reasonable to briefly discuss the
values of these parameters. For instance the value
$\lambda_s/l = 100$ means that $\lambda_s = 1\mu \mathrm{m}$ at
$l = 10 \, \mathrm{nm}$. As far as the values of the parameter $d_d/l$
are concerned, it is difficult \cite{Chung02} to evaluate the kinetic
length $d_d$. We should have in mind, however, that the
Eqs.~(\ref{Gl10a},\ref{Gl10b}) are valid under the assumption
$(x_i - x_{i-1}) \ll d_d$ and, therefore, we have to take
$d_d/l \gg 1$. The value of the parameter $l_0/l$
was assumed to be $l_0/l = 0.24$ in order to keep
the interstep distance in the bunch to be in a convenient interval. 
Finally, the values of $\beta$ used were $\beta = 0.01$ and $\beta = 0.1$ 
corresponding to a rather high Ehrlich-Schwoebel barrier.
Figure \ref{2} shows four sets of data for $l_{\mathrm{min}}$ as
a function of $N$ obtained for two different step interaction laws, $n=2$
and $n=3$, using the natural bunching as well as the single bunch
initial conditions. In all cases excellent agreement with the 
theoretical prediction (\ref{lminNgen}) is found. 
The same quality of agreement has been obtained for the problem of
growth with inverse Ehrlich-Schwoebel barriers \cite{Tonchev03}.

\begin{figure}
\centerline{\epsfig{file=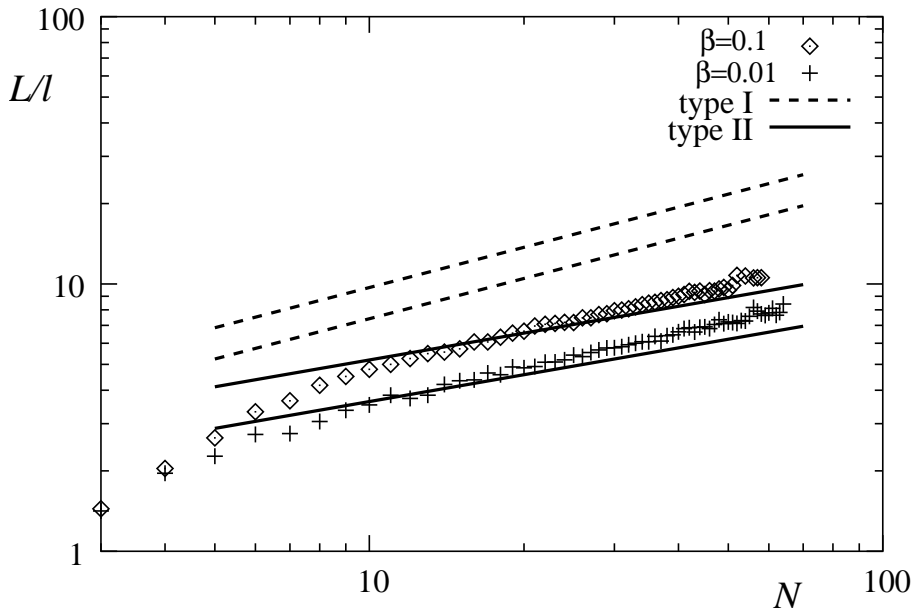,width=12.cm}}
\vspace*{0.5cm}
\caption{\label{3} Numerical results for the bunch width $L$ as a 
function of bunch size. Data are shown for two of the 
parameter sets displayed in Fig.\ref{2}: 
$\beta = 0.1$, $\lambda_s/l = 200$, $d/l = 150$ (diamonds), and
$\beta = 0.01$, $\lambda_s/l = 100$, $d/l = 10$ (crosses). In both cases
$l_0/l = 0.24$ and the system contained 500 steps. Dashed and full
lines show the predictions for type I and type II solutions,
respectively.  A power law fit to the data yields
$L \sim N^{0.44}$.} 
\end{figure}

\begin{figure}
\centerline{\epsfig{file=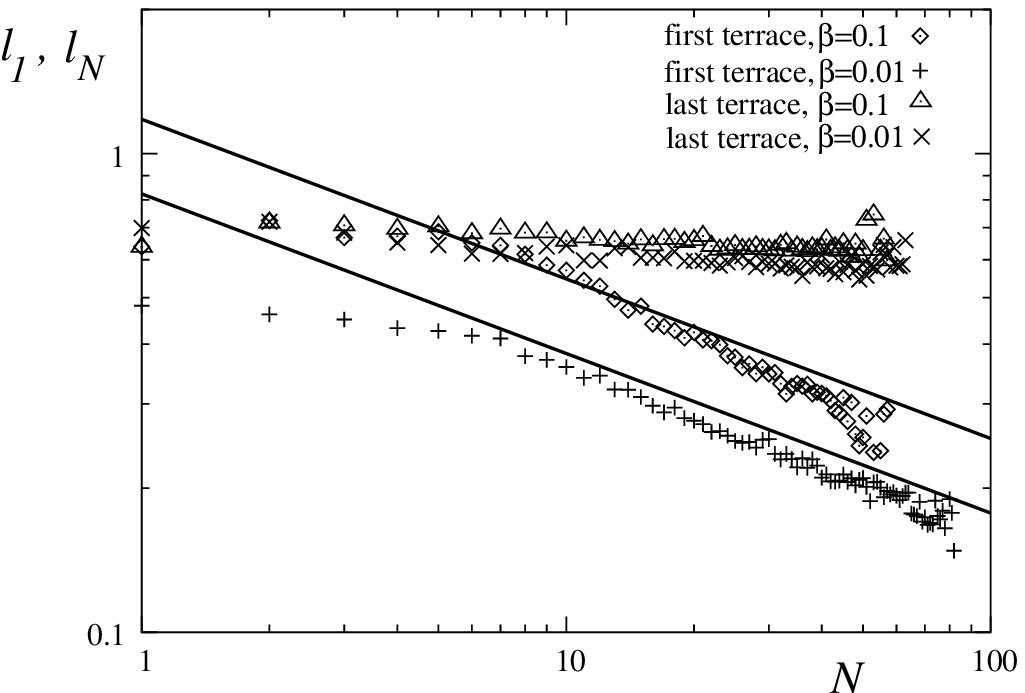,width=12.cm}}
\vspace*{0.5cm}
\caption{\label{4} Scaling of the first and last terrace size in the
bunch with bunch size for the same parameter sets as in Fig.\ref{3}.
The full lines show the theoretical prediction for type II solutions.
}
\end{figure}

In Fig.\ref{3} we show data for the dependence of the total bunch
width $L$ on the number of steps. 
Although the overall magnitude of $L$ is consistent with the prediction 
(\ref{LN}) for type II stationary profiles (and rules out
type I behavior), a power law fit to the
data yields an estimate $\alpha \approx 0.44$ which is intermediate
between the type II ($\alpha = 1/3$) and type I ($\alpha = 1/2$) 
predictions. 
To gain some insight into this discrepancy, we take a closer
look at the shapes of the bunches in the numerical simulation (Fig.\ref{1}).
It is clear that the bunches are distinctly \textit{asymmetric}: While
there is an abrupt change in the terrace length at the lower (trailing)
edge of the bunch, at the upper (leading) edge the terrace lengths increase
gradually. The asymmetry can be quantified by looking at the scaling of 
the size of the first ($l_1$) and last ($l_N$) terrace in the bunch
with the bunch size $N$ (Fig.\ref{4}). While the data for $l_1$
are in good agreement with the theoretical prediction (\ref{lminNgen}),
the size of the last terrace is found to be essentially independent of 
$N$. Incidentally, the latter behavior is also characteristic of the type I
stationary profiles (see Sect.\ref{Scaling}). More significantly, a constant
last terrace size $l_N \sim l$ results trivially from our way of numerically locating
the bunch edge, if the terrace size increases continuously across
the mean terrace size as one moves out of the bunch in the forward direction,
i.e., if a sharply defined bunch edge in fact does not exist. 

What is the origin of the asymmetry in the bunch shape? We have shown in 
Sect.\ref{Corrections} that the symmetry-breaking term causes the bunch
edge to fray out at one side, in a qualitatively similar manner
to the behavior seen in Fig.\ref{1}. However, the blurring
of the bunch edge is predicted to occur at the lower (trailing) edge, rather
than at the upper edge, and in addition the effect becomes negligibly small
already for moderately sized bunches (see Fig.\ref{Fig2}). We believe, instead, that the 
bunch asymmetry is intimately related to the exchange of \textit{crossing steps} between bunches.
These steps gradually accelerate out of the bunch at the leading edge, which
translates into a gradual increase of the mean terrace size. Conversely, when
approaching the next bunch from behind, the crossing step decelerates quite
abruptly, because it is primarily fed from behind (this is particularly true
in the almost one-sided regime mainly considered in our simulations). 
The exchange of steps between bunches also implies that the bunches
move laterally at a speed which is different from the mean sublimation
rate. To capture
the asymmetry in the continuum theory, it will therefore be necessary to go
beyond the stationary solutions considered in Sect.\ref{Continuum}, which by
construction are symmetric, and to investigate solutions describing moving
and interacting bunches. It is worth pointing out that the existence 
of crossing steps partly invalidates the argument used in Sect.\ref{Current} 
to fix the mean surface current, because the argument assumes that all
steps reside in bunches \cite{Liu98a}.

We close this section with some remarks concerning the limiting case (a) of 
Section \ref{Equationsofmotion}.   
We have shown that no bunching occurs in this case,
i.~e., when the assumption $x_i - x_{i-1} > d_d$ is fulfilled, if only
the linear and quadratic terms [$(x_i-x_{i-1})/\lambda_s$
and $(x_i-x_{i-1})^2/\lambda_s^2$] in the
expressions for the step velocity are taken into account. To clarify the
problem of step bunching instability in the limiting case (a) we
did a lot of numerical work making use of the full expressions for 
the step velocity. Integration of the equations of step motion
proved the existence of step bunching at parameter values $\beta = 0.01,\ 
l_0/l = 0.003$ and $d_d/l = 1/3, d_d/l = 1/30$ and $d_d/l = 1/300$. It
is essential to note, however, that the magnitude of the step-step
repulsion energy used in these integration runs was much smaller
than in the integration of the equations obtained in the case
$d_d \gg l$, where we used $l_0/l = 0.24$. On the other hand, the density of steps
in the bunch is higher in the case $d_d \gg l$ compared with the
opposite limiting case $d_d \ll l$. These findings indicate a
strong impact of the parameter $d_d$ on the bunching process.
When the parameter $d_d$ is larger than the interstep distance
$l$, bunching occurs even at a very strong repulsion between
the steps and the step density in the bunch
is rather high. On the contrary, when the parameter $d_d$ is
smaller than the interstep distance $l$, bunching occurs only
at a very weak repulsion between the steps and the step density
in the bunch is relatively small. This is not surprising because
in the case under consideration neither the linear nor the quadratic
term induces instability of the vicinal surface. Destablising
terms are of higher order, i.~e.\ 
$[(x_i - x_{i-1})/\lambda_s]^\nu$ with $\nu > 2$ and
their effect is relatively weak so that it cannot dominate a
strong repulsion between the steps. It is interesting to note,
however, that in this case of weak instability ($d_d \ll l$)
the minimum interstep distance in the bunch scales with the
number of steps in exactly the same way $(l_{\min} \sim N^{-2/3})$
as in the case of strong instability ($d_d \gg l$). Further discussion
of this regime will be presented elsewhere.

\section{Universality classes of step bunching}
\label{Universality}

Before drawing some general conclusions in the next section, here
we wish to put our work into the context of a classification scheme
for step bunching instabilities proposed by Pimpinelli, Tonchev, Videcoq and
Vladimirova (PTVV).  
\cite{Pimpinelli02}. It is based on a generic continuum equation of the form
\be
\label{PimpEq}
\frac{\partial h}{\partial t} + \frac{\partial}{\partial x} 
\left[ K_1 m^\varrho + \frac{K_2}{m^k} \frac{\partial^2}{\partial x^2}
m^n \right] = {\mathrm{const.}}
\ee
Here $K_1$ and $K_2$ are material constants with $K_2 > 0$ and $K_1 \varrho > 0$, the slope
$m = \partial h/\partial x$ is assumed positive everywhere, 
and $\varrho$, $k$ and $n$ are 
exponents characterizing a class of step bunching instabilities. The exponent
$n$ is simply the exponent of the repulsive step interaction, and the exponent
$k$ reflects the slope dependence of the surface mobility; $k=1$ and $k=0$
correspond to slow and fast detachment/attachment kinetics, respectively 
(see Sect.\ref{Contrelax}). Equation (\ref{PimpEq}) is a slight generalization
\cite{Krug04a} of the equation proposed in Ref.36, where only the case $k=0$
was considered.   

PTVV argued that the characteristic scaling exponents $\alpha$ and $z$
introduced in Sect.\ref{Introduction} can be extracted from (\ref{PimpEq}) by
requiring that the equation should be invariant under the scale transformation
\be
\label{scale}
h(x,t) \Rightarrow b^{-\alpha} h(bx, b^z t)
\ee
for an arbitrary scale factor $b$. This yields the expressions 
\be
\label{exponents}
\alpha = 1 + \frac{2}{n-k-\varrho}, \;\; \gamma = \frac{2}{2 + n - k - \varrho}, \;\;
z = \frac{2(1 + n - k - 2\rho)}{n - k - \varrho},
\ee
where the scaling relation (\ref{scalingrelation}) has been used. 

Apart from the symmetry-breaking term in (\ref{current}), 
the continuum equation (\ref{full}) for the sublimation problem
is of the generic form (\ref{PimpEq}) with $\varrho = -1$ and $k=1$. 
It is straightforward to check that, under the rescaling (\ref{scale}), 
the symmetry-breaking term is
smaller by a factor of $b^{-\alpha}$ compared to the leading term
$\sim 1/m$, and hence it becomes negligible at large scales; this is consistent
with the detailed analysis in Sect.\ref{Corrections}.  
For $\varrho = -1$ and $k=1$ the exponents 
in (\ref{exponents}) reduce to $\alpha = (n+2)/n$ and $\gamma = 2/(2+n)$,
which we readily recognize as the scaling exponents characteristic
of \textit{type I} solutions [compare to (\ref{ScalinggenI})]. To obtain
the exponents for type II solutions (which, as was shown in Sect.\ref{Numerics}, 
correctly describe the bunch shape), we have
to set $\varrho = 0$ instead of $\varrho = -1$ in (\ref{exponents}). 

The reason for the shift in the value of $\varrho$ is evident in view of the  
considerations of Sect.\ref{Mechanics}. The scaling argument of PTVV assumes
that all terms in (\ref{PimpEq}) are of a similar order of magnitude; in particular,
the stabilizing term is balanced against the destabilizing current $B m^\varrho$.
However, for type II solutions the total current is fixed at a value $J_0$
which is independent of the slope, and which dominates over the term $B m^\varrho$
for large slopes when $\varrho < 0$. Thus the stabilizing term is balanced
against a \textit{slope-independent} current, which effectively implies that 
$\varrho = 0$. 

The argument clearly extends to any negative value of $\varrho$,
and suggests that generally $\varrho$ should be replaced by $\max(0,\varrho)$.
For the continuum evolution equation (\ref{rho=1equation}) describing
electromigration-induced step bunching in the diffusion-limited regime,
which corresponds to $\varrho = 1$ and $k=0$, the ambiguity regarding type I
and type II solutions does not arise (see Sect.\ref{Electromigration}). Since the 
static exponents $\alpha$ and $\gamma$ in (\ref{exponents}) 
only depend on the sum $\varrho + k$, it is evident that they take the 
same values for $\varrho = 0$, $k=1$ as for $\varrho = 1$, $k=0$; for this
reason the static scaling exponents for electromigration-induced step bunching
take the same values in the diffusion-limited and the attachment/detachment-limited
regimes.

\begin{figure}
\centerline{\epsfig{file=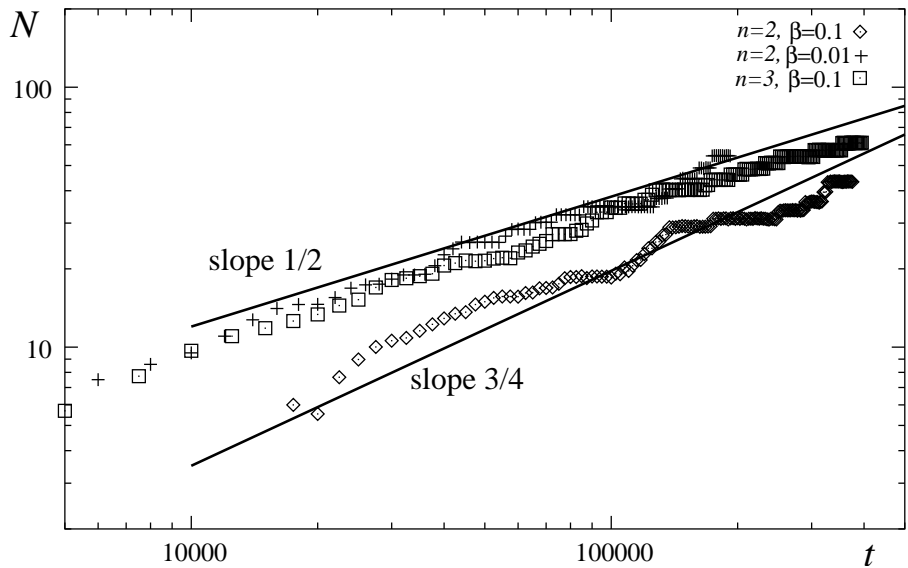,width=12.cm}}
\vspace*{0.5cm}
\caption{\label{5} Time dependence of the mean bunch size for 
the parameter sets in Figs.\ref{3} and \ref{4}, and an additional
set with $\beta = 0.1$, $\lambda_s/l = 100$, $d/l = 100$, $l_0/l = 
0.24$, and step interaction exponent $n=3$. The bold lines illustrate
the predictions of the scaling theory for $\varrho = -1$ ($\alpha/z = 1/2$)
and for $\varrho = 0$, $n=2$ ($\alpha/z = 3/4$).
}
\end{figure}

The scaling theory of PTVV also makes predictions about the coarsening behavior
of the bunched surface, which our analysis of stationary bunch shapes clearly
cannot address. Setting $\varrho = -1$, the expressions (\ref{exponents}) yield 
$\alpha/z = 1/2$ for the exponent in the coarsening law (\ref{DynExp}), which is 
\textit{independent} of both $n$ and $k$; on the other hand, with $\varrho = 0$
one obtains $\alpha/z = (n+1)/2n$ for $k=1$. In Fig.\ref{5} we compare
numerical data for the temporal evolution of the mean bunch size to these
two coarsening laws. The simulations seem consistent with the ``superuniversal''
value $\alpha/z = 1/2$, but also $\alpha/z = 3/4$ (for $n=2$) or 2/3 (for $n=3$)
cannot be ruled out. More extensive simulations are needed to firmly
pin down the coarsening behavior; this is particularly true here because, in contrast
to the static scaling properties discussed earlier in Sect.\ref{Numerics}, we do
not have any analytic information about the coefficient of the coarsening law. 
In a recent study of a simple toy model of step bunching, which ignores
the repulsive step-step interactions and allows steps to coalesce, it
was necessary to go to extremely long times, equivalent to the growth of
more than $10^5$ monolayer, to ascertain the true asymptotic coarsening behavior \cite{Slanina04}.

\section{Conclusions}
\label{Conclusions}

In this paper we have presented a detailed analysis of the step bunching
instability caused by an Ehrlich-Schwoebel effect during sublimation
in the limit of a small desorption rate.
When the kinetic length $d_d = D_s/K_d$ is large compared to the 
average distance between the steps, the
instability is strong and bunches of steps appear even at strong
repulsion between the steps. In the
opposite case (kinetic length $d_d$ smaller than the
interstep distance) the instability is weak and bunches occur
only when the step-step repulsion is several orders of magnitude
weaker than in the previous case. 

A central part of the work is the derivation of the continuum evolution equation 
in Section \ref{Continuum}, and the careful analysis of its stationary
bunch solutions. We have shown that two different types of stationary
solutions with different scaling properties can be found, depending
on whether the mean surface current $J_0$ is kept fixed or not. Following
Ref.24, we have argued that $J_0$ is independent of bunch size,
and that the correct bunch shape is given by the type II solution, which
describes a bunch of finite extent with Pokrovsky-Talapov-type singularities
at the edges. This is confirmed by the excellent agreement with numerical
simulation results for the minimal interstep spacing $l_{\mathrm{min}}$ and
the first interstep spacing in the bunch $l_1$ presented in Sect.\ref{Numerics}.

On the other hand, we find noticeable deviations of the behavior of the
total bunch width $L$ from the type II prediction. 
We suggest that the discrepancy may 
be related to the distinct asymmetry between the leading and the trailing edges
of the bunch: The terraces between the crossing 
steps escaping from the leading edge of the bunch appear to contribute
strongly to the total bunch width, to the extent that asymptotically
$L$ may be considerably larger than 
$N l_{\mathrm{min}}$. Further clarification of the issue requires
a better understanding of the motion of bunches and the interactions between
bunches, which is beyond the scope of the present paper.
 
Our work has important consequences for the recently proposed scenario of
universality classes for step bunching instabilities \cite{Pimpinelli02}.  
First, we have pointed out the mathematical equivalence between 
appropriate limits of the step bunching instabilities caused
by sublimation, growth and electromigration on the levels of \textit{both} 
discrete step dynamics and continuum evolution equations. This equivalence gives
a very clear meaning to the notion of a universality class, and we believe
that the particular class considered in this paper (characterized, in essence,
by the linearity of the destabilizing terms in the step equations)
is in fact relevant to a wide range of experimentally realized systems.
As a practical matter, the unified view provided by the continuum approach
allows us to derive explicit formulae for the bunch shape which, through
the identification of the scaling parameter $S$, are directly applicable
to these diverse realizations of step bunching.    
Second, we have shown that the procedure employed in Ref.36
to extract the scaling exponents from the continuum evolution equation captures
only one type of solution (the type I solutions of Sect.\ref{Mechanics}), which
is not the relevant one at least as far as the time-independent scaling
properties are concerned.

A crucial question that should be addressed in future work concerns the 
coarsening behavior of the bunched surface, and the relationship between
coarsening dynamics and bunch motion. As was discussed in Sect.\ref{Universality},
the present work remains inconclusive on this point. It is remarkable, however,
that a very robust scaling of the mean bunch size and bunch spacing as $N \sim \xi \sim t^{1/2}$
has been observed in a number of numerical simulations, both for electromigration
\cite{Dobbs96,Liu98a,Sato99} 
and growth with inverse ES barriers \cite{Sato01,Tonchev03}, 
as well as in an experimental study of 
electromigration-induced step bunching on Si(111) \cite{Yang96}. 
Liu and Weeks \cite{Liu98a} have proposed an elegant explanation for the ubiquity of the 
$t^{1/2}$-scaling within a continuum setting; their argument presupposes, however,
as does the scaling approach of PTVV \cite{Pimpinelli02},
that the bunch spacing $\xi$ is the only lateral length scale in the problem,
although the bunch width $L$ clearly comes into play as well \cite{Krug04a}. 
This remains true even if the internal bunch structure is eliminated by allowing
the steps to coalesce \cite{Slanina04}. 
Thus the origin of the observed temporal scaling remains to be understood.

\section*{Acknowledgements} 

This work has been supported by DFG within SFB 616 \textit{Energiedissipation 
an Oberfl\"achen}. 
V. Tonchev acknowledges the financial support from EC RTD PROJECT
IST-32264-2001 - Nanocold, NTT Basic Research Labs through the
\textit{Visiting Researchers Program},  and NATO Science for Peace
Project 977986. J.K. wishes to thank V. Popkov for useful discussions.


\begin{thebibliography}{99}

\bibitem{Venezuela99} P.~Venezuela, J.~Tersoff, J.~A.~Floro, E.~Chason,
                        D.~M.~Follstaedt, Feng Liu, and M.~G.~Lagally, Nature
                        {\bf 397}, 678 (1999).
\bibitem{Teichert02} C.~Teichert, Phys. Rep. {\bf 365}, 335 (2002).
\bibitem{Neel03} N. N\'eel, T. Maroutian, L. Douillard, and H.-J. Ernst, 
J. Phys.: Condens. Matter {\bf 15}, S3227 (2003).
\bibitem{Syvajarvi02} M. Syv\"aj\"arvi, R. Yakimova, and E. Janz\'en, J. Cryst. Growth ,
\textbf{236}, 297 (2002).
\bibitem{Gliko02} O. Gliko, N. Booth and P.G. Vekilov, Acta Cryst. \textbf{D58}, 1622 (2002).
\bibitem{Gliko03} O. Gliko, I. Reviakine and P.G. Vekilov, Phys. Rev. Lett. \textbf{90}, 225503 (2003).
\bibitem{Tersoff95} J.~Tersoff, Y.H.~Phang, Z.~Zhang, and M.G.~Lagally,
                    Phys.~Rev.~Lett.~{\bf 75}, 2730 (1995).
\bibitem{Liu98} F.~Liu, J.~Tersoff, and M.G.~Lagally, Phys.~Rev.~Lett.~
                  {\bf 80}, 1268 (1998).
\bibitem{Schwoebel69} R.~L.~Schwoebel, J.~Appl.~Phys. {\bf 40}, 614 (1969).
\bibitem{Pimpinelli94} A.~Pimpinelli, I.~Elkinani, A.~Karma, C.~Misbah,
                          and J.~Villain, J.~Phys.: Condens. Matter {\bf 6}, 2661 (1994).
\bibitem{Uwaha95} M.~Uwaha, Y.~Saito, and M.~Sato, J.~Chryst.~Growth
                      {\bf 146}, 164 (1995).
\bibitem{Sato01} M. Sato and M. Uwaha, Surf. Sci. \textbf{493}, 494 (2001).
\bibitem{Xie02} M.H. Xie, S.Y. Leung, and S.Y. Tong, Surf. Sci. \textbf{515}, L459 (2002).
\bibitem{Tonchev03} V. Tonchev, J. Krug, H. Omi, Y. Homma, and A. Pimpinelli (manuscript in preparation).
\bibitem{Latyshev89} A.~V.~Latyshev, A.~L.~Aseev, A.~B.~Krasilnikov, and
                       S.~I.~Stenin, Surf.~Sci.~{\bf 213}, 157 (1989).
\bibitem{Homma90} Y.~Homma, R.~J.~Mcclelland, and H.~Hibino,
                    Jpn.~J.~Appl.~Phys.~{\bf 29} L2254 (1990).
\bibitem{Stoyanov91} S.~Stoyanov, Jpn.~J.~Appl.~Phys.\ {\bf 30}, 1 (1991).
\bibitem{Houchmandzadeh94} B.~Houchmandzadeh, C.~Misbah, and A.~Pimpinelli,
                              J.~Phys.~I France {\bf 4}, 1843 (1994).
\bibitem{Yang96} Y.-N. Yang, E.S. Fu, and E.D. Williams, Surf. Sci. \textbf{356}, 101 (1996).
\bibitem{Dobbs96} H. Dobbs and J. Krug, J. Phys. I France \textbf{6}, 413 (1996).
\bibitem{Stoyanov97} S.~Stoyanov, in: M.~Tringides (Ed.), \textit{Surface
                        Diffusion -- Atomistic and Collective Processes}, Plenum,
                        New York, 1997, p.~285.
\bibitem{Stoyanov97a} S.~Stoyanov, Surf. Sci.\ {\bf 370}, 345 (1997).
\bibitem{Fu97} E.~S.~Fu, D.-J.~Liu, M.~D.~Johnson, J.~D.~Weeks, and
                  E.D.~Williams, Surf.~Sci.\ {\bf 385}, 259 (1997).
\bibitem{Liu98a} D.-J. Liu and J.D. Weeks, Phys. Rev. B \textbf{57}, 14891 (1998).
\bibitem{Metois99} J.~J.~M\'etois and S.~Stoyanov, Surf.~Sci.\ {\bf 440}, 407 (1999).
\bibitem{Sato99} M.~Sato and M.~Uwaha, Surface Sci. {\bf 442}, 318 (1999).
\bibitem{Yagi01} K. Yagi, H. Minoda, and M. Degawa, Surf. Sci. Rep. {\bf 43}, 45 (2001).
\bibitem{Minoda03} H. Minoda, J. Phys.: Condens. Matter {\bf 15}, S3255 (2003).
\bibitem{Stoyanov98a} S.~Stoyanov and V.~Tonchev, Phys.~Rev. B {\bf 58}, 1590 (1998).
\bibitem{Stoyanov98b} S.~Stoyanov, Surf. Sci. {\bf 416}, 200 (1998).
\bibitem{Fujita99} K.~Fujita, M.~Ichikawa, and S.S.~Stoyanov, Phys.~Rev. B {\bf 60}, 16006 (1999).
\bibitem{Homma00} Y. Homma and N. Aizawa, Phys. Rev. B \textbf{62}, 8323 (2000).
\bibitem{Stoyanov00} S.~Stoyanov, J.~J.~M\'etois, and V.~Tonchev, Surf.~Sci.\ {\bf 465}, 227 (2000).
\bibitem{Stoyanov00a} S. Stoyanov, Surf. Sci. \textbf{464}, L715 (2000)
\bibitem{Nozieres91} P. Nozi\`eres, in C. Godr\`eche (Ed.), \textit{Solids Far From Equilibrium}, 
Cambridge University Press, Cambridge 1991.
\bibitem{Pimpinelli02} A. Pimpinelli, V. Tonchev, A. Videcoq, and M. Vladimirova,
Phys. Rev. Lett. \textbf{88}, 206103 (2002).
\bibitem{Burton51} W. Burton, N. Carbrera, and F. C. Frank, Trans. Roy. Soc. A \textbf{243}, 299 (1951)
\bibitem{Krug04b} J. Krug, in A. Voigt (Ed.), \textit{Multiscale Modeling of 
Epitaxial Growth} (Birkh\"auser, 2004) ({\tt{cond-mat/0405066}}).
\bibitem{Krug04a} J. Krug, in G. Radons, P. H\"aussler, W. Just (Eds.),
\textit{Collective Dynamics of Nonlinear and Disordered Systems} (Springer, Berlin 2004).
\bibitem{Michely03} T. Michely, J. Krug: \textit{Islands, Mounds and Atoms. Patterns and
Processes in Crystal Growth Far from Equilibrium} (Springer, Berlin 2004).
\bibitem{Chung02} W.F. Chung and M.S. Altman, Phys. Rev. B \textbf{66}, 075338 (2002).
\bibitem{Krug97} J.~Krug, in D.~Kim et al. (Eds.), \textit{Dynamics of fluctuating
                    interfaces and related phenomena}  (World Scientific, Singapore 1997), p. 95.
\bibitem{Frank62} F.~C.~Frank, \textit{Growth and perfection of Crystals},
                     edited by R.~Doremus, B.~Roberts and D.~Turnbull
                     (John Wiley \& Sons, New York, 1962) p.~411.
\bibitem{Nozieres87} P. Nozi\`eres, J. Physique \textbf{48}, 1605 (1987).
\bibitem{Liu96} D.-J.~Liu, E.S.~Fu, M.D.~Johnson, J.D.~Weeks, and E.D.~Williams,
                   J.~Vac.~Sci.~Technol.~B {\bf 14}, 2799 (1996).
\bibitem{Lancon90} F.~Lan\c{c}on and J.~Villain, in: M.~Lagally (Ed), \textit{Kinetics
                      of ordering and growth at surfaces}, Plenum, New York,
                      1990.
\bibitem{Mullins59} W.~W.~Mullins, J.~Appl.~Phys.~{\bf 30}, 77 (1959).
\bibitem{Politi96} P. Politi and J. Villain, Phys. Rev. B \textbf{54}, 5114 (1996).
\bibitem{Villain91} J. Villain, J. Phys. I (France) \textbf{1}, 19 (1991).
\bibitem{Krug93} J. Krug, M. Plischke, and M. Siegert, Phys. Rev. Lett. \textbf{70}, 3271 (1993)
\bibitem{Politi00} P. Politi, G. Grenet, A. Marty, A. Ponchet, and J. Villain: Phys. Rep. \textbf{324},
271 (2000).
\bibitem{Myslivecek02} J. Myslive\v{c}ek,  C. Schelling,
F. Sch\"{a}ffler, G. Springholz, P. \v{S}milauer, J. Krug, and 
B. Voigtl\"{a}nder, Surf. Sci. \textbf{520}, 193 (2002).
\bibitem{Slanina04} F. Slanina, J. Krug, and M. Kotrla (manuscript in preparation). 

\end{thebibliography}
\end{document}